\documentclass{article}
\usepackage[english]{babel}
\usepackage{amsmath,amssymb,graphicx,xcolor,latexsym,theorem}
\usepackage{comment}
\usepackage{booktabs}
\usepackage{amsmath}
\usepackage{subcaption}
\usepackage{bbm}
\newcommand{\ra}[1]{\renewcommand{\arraystretch}{#1}}

\newcommand{\ip}[2]{\big \langle #1,#2 \big \rangle}

\newcommand{\vertiii}[1]{{\left\vert\kern-0.25ex\left\vert\kern-0.25ex\left\vert #1 
    \right\vert\kern-0.25ex\right\vert\kern-0.25ex\right\vert}}
\newenvironment{proof}{\noindent\textbf{Proof\ }}{\hspace*{\fill}$\Box$\medskip}

{\theorembodyfont{\rmfamily}\newtheorem{example}{Example}}

\newtheorem{theorem}{Theorem}[section]
\newtheorem{remark}[theorem]{Remark}

\newtheorem{proposition}[theorem]{Proposition}

\theoremstyle{break}
\newcommand{\R}{\mathbb{R}}

\newcommand{\argmin}{\mathop{\mathrm{argmin}}}

\DeclareMathOperator*{\esssup}{ess\,sup}

\usepackage[colorlinks,citecolor=red,urlcolor=blue,bookmarks=false,hypertexnames=true]{hyperref}
\usepackage{shuffle}
\usepackage{mathrsfs}
\usepackage{mathtools}
\usepackage{atbegshi}
\usepackage[colorlinks]{hyperref}
\usepackage[nameinlink,capitalize]{cleveref} 
\usepackage[autostyle]{csquotes}
\usepackage[toc,page]{appendix}

\usepackage{todonotes}
\begin{document}
\title{Pricing American options under rough volatility using deep-signatures and signature-kernels}

\author{Christian Bayer\thanks{Weierstrass Institut (WIAS), Berlin, Germany, email: bayerc@wias-berlin.de.} \quad Luca Pelizzari\thanks{Technische Universität Berlin and Weierstrass Institut (WIAS), Berlin, Germany, email: pelizzari@wias-berlin.de} \quad Jia-Jie Zhu \thanks{Weierstrass Institut (WIAS), Berlin, Germany, email: zhu@wias-berlin.de.}}
\maketitle
\begin{abstract} We extend the signature-based primal and dual solutions to the optimal stopping problem recently introduced in [Bayer et al.: Primal and dual optimal stopping with signatures, to appear in Finance \& Stochastics 2025], by integrating deep-signature and signature-kernel learning methodologies. These approaches are designed for non-Markovian frameworks, in particular enabling the pricing of American options under rough volatility. We demonstrate and compare the performance within the popular rough Heston and rough Bergomi models.  \\ \\ \textbf{Keywords:} Signature, optimal stopping, rough volatility, deep learning, kernel learning. \\ \textbf{MSC2020 classifications:} 60G40, 60L10, 91G20, 91G60.
\end{abstract}

\tableofcontents

\section{Introduction}

Starting with the seminal paper \cite{kalsi2020optimal}, (rough) path signatures \cite{chen1957integration} have been increasingly recognized as a powerful tool for numerical approximations to solutions of stochastic optimal control problems when the underlying system does not have the Markov property.
While the workhorse theoretical methods for stochastic optimal control -- dynamic programming / HJB equations on the one hand, Pontryagin maximum principle / BSDEs on the other hand -- can, in principle, be formulated and analysed even when the underlying system is not Markovian, this comes at the expense of tractability.

For example, in the Markovian case, we can assume (under mild conditions) that optimal controls are of feedback form, i.e., can be expressed as functions $\alpha^\ast(t, X_t)$ of the underlying state variable $X_t$.
If we do not make the Markovian assumption, we only know that optimal controls $\alpha^\ast_t$ are, say, progressively measurable w.r.t.~the governing filtration, i.e., $\alpha^\ast_t = \alpha^\ast\left(t, X|_{[0,t]}\right)$, provided that said filtration is generated by a process $X$.
Similar remarks can be made for the value function as well as the solutions to the corresponding BSDE systems.

From a numerical point of view, this observation increases the computational challenge considerably for solving non-Markovian stochastic optimal control problems.
In essence, an approximation problem for a finite-dimensional function, say, $\alpha^\ast : [0,T] \times \R^d \to \R^m$, is replaced with one for a function on pathspace, say $\alpha^\ast : [0,T] \times C([0,T]; \R^d) \to \R^m$ (with additional measurability constraints).

Kalsi, Lyons and Perez Arribas suggested a general framework for solving stochastic optimal control problems in a model-free way (in particular, without assuming a Markovian problem) for the example of an optimal execution problem, see \cite{kalsi2020optimal}.
In a nutshell, the approach consists of approximating the strategy as well as the value function as linear functionals of the signature of the underlying process.
The \emph{linearization} -- based on the signature's \emph{universality} -- allows them to rephrase the optimal execution problem as a \emph{deterministic} optimization problem in terms of the \emph{expected signature}.

Full linearization as in \cite{kalsi2020optimal} is only feasible if the unknown function (value function, control, \ldots) is smooth enough, so as not to require a very deep degree of signature approximation -- comparable to approximations with polynomials of high degree in finite dimensional spaces.
Otherwise, several other strategies have been considered in the literature:
\begin{itemize}
    \item Signatures can be interlaced with non-linear transformations of the path, e.g., deep neural networks, to increase the signature's ``expressivity'', see, for instance, \cite{toth21}.
    \item Signatures -- or, alternatively, log-signatures to reduce the dimension -- can be used as features for other, non-linear approximation methods, e.g., deep neural networks. In the context of optimal stopping, this was first suggested in \cite{bayer2021optimal}.
\end{itemize}

In addition, \emph{signature kernels}, see \cite{chevyrev2018signature}, provide a classical kernel learning approach to regression problems in general, and can also be used for stochastic optimal control, in particular.
\cite{salvi2021signature} observed that the (non-truncated) signature kernel satisfies a Goursat PDE, and, hence, can be computed numerically, without requiring the (otherwise crucial) truncations of the signature. In the context of optimal stopping, \cite{horvath2023optimal} exploit this kernel-trick by introducing a \emph{distributional regression} approach to learn the optimal stopping value, applying kernel-ridge regression based on the \emph{higher-order signature kernel}.

\emph{Rough volatility models}, see \cite{bayer2023rough}, are a popular class of stochastic volatility models in finance, i.e., with a stock price process following a dynamics $dS_t = \sqrt{v_t} S_t dZ_t$, where the stochastic variance process $v$ is ``rough'', e.g., an exponential of fractional Brownian motion with Hurst index $0 < H < 1/2$.
Crucially, such models $(S,v)$ do not have the Markov property, so signature methods are ideally suited for computing approximate solutions to the optimal stopping problem in $S$ -- in financial terms: to compute the prices of \emph{American or Bermudan} options.

This paper builds on \cite{bayer2023primal}, where adaptations of the classical Longstaff--Schwartz and dual algorithms for Bermudan option pricing based on linear functionals of the signature were introduced and analysed.
In this paper, we further extend these algorithms by introducing versions based on non-linear functions of the signature as well as signature-kernel versions.
We then apply them to the Bermudan options pricing problem for popular rough volatility models (specifically, the rough Bergomi model \cite{bayer2016pricing} and the rough Heston model \cite{el2019characteristic}), and numerically analyse their performance for these models under realistic model parameters.

The goal of this paper is to compute American (more precisely, Bermudan) option prices in the aforementioned rough volatility models, with special emphasis on numerical performance as well as comparisons between the methods outlined before.
Specifically, we solve primal and dual formulations of the optimal stopping problem -- giving us lower and upper bounds of the option price, respectively -- using three different signature methods each:
\begin{enumerate}
    \item linear functionals of the truncated signature;
    \item deep neural networks applied to truncated signatures;
    \item linear combinations of the signature kernel.
\end{enumerate}
Compared to the literature (see, for instance, \cite{bayer2023primal} based on linear functionals of the signature alone), we provide considerably sharper bounds for the option price, in the sense of the length of the interval between lower and upper bounds.
More specifically, we find that for a realistic rough volatility model (the rough Bergomi model with Hurst index $H=0.07$, as suggested in \cite{bayer2016pricing}), we obtain a tight gap of about $1\%$.

Regarding the comparison between the different methods, no method seems to consistently outperform the others.
For the primal problem, all three methods tend to perform very well, with mild advantages for the linear and the signature kernel methods.
The dual formulation, however, seems to lead to more difficult approximation and optimization problems, and the neural-network-based method has some advantages.
Overall, we see the biggest improvements compared to \cite{bayer2023primal} for the dual formulation.

This conclusion is also supported by various other numerical studies performed, including a study of the dependence of the training error on the number of training samples.
We also study the relative importance of the different components in the signature, specifically in the DNN-based method.
For the primal formulation, while the most important signature components are those corresponding to powers of the increments, generally all components matter, and there is no apparent sparse structure.
On the other hand, for the dual problem, only very few signature components actually are important, and there seems to be great potential for dimension reduction.

\paragraph{Acknowledgments}
All authors gratefully acknowledge funding by the Deutsche Forschungsgemeinschaft (DFG, German Research Foundation) under Germany’s Excellence Strategy – The Berlin Mathematics Research Center MATH+ (EXC-2046/1, project ID: 390685689). CB and PL acknowledge support from DFG CRC/TRR 388 ``Rough Analysis, Stochastic Dynamics and Related Fields'', Project B03. We would like to thank Wenyuan Li for identifying errors in the European prices in earlier versions of Tables \ref{tab:putIntervals_08_bergomi} and \ref{tab:putIntervals_H07_Bergomi}.

\section{A review on Monte-Carlo methods for optimal stopping}\label{sec:Optimal stopping}
In this section we introduce the optimal stopping problem in a fairly general framework, and recall two general simulation based procedures to derive lower and upper bounds of the optimal value. Let $\{\Omega, \mathcal{F}, \mathbb{P}\}$ be a probability space supporting a state-process $(X_t:0\leq t \leq T)$, which we only assume to be $\alpha$-Hölder continuous almost surely, $\alpha \in (0,1)$. Moreover, we consider an $(\mathcal{F}^X_t)$-adapted, continuous process $(Z_t:0\leq t \leq T)$, the cash-flow process, which for technical reasons we assume to fulfill $\sup_{0\leq t \leq T}|Z_t|^2 \in L^1$. Finally, the optimal stopping problem reads \begin{equation}\label{eq:optimalstopping}
Y_0 = \sup_{\tau \in \mathcal{S}_0}\mathbb{E}[Z_\tau], \quad Y_t =\esssup_{\tau \in \mathcal{S}_t}\mathbb{E}[Z_{\tau}|\mathcal{F}^X_t], \quad 0 < t \leq T,
\end{equation}where $\mathcal{S}_t$ denotes the set of $(\mathcal{F}^X_t)$-stopping times on $[t,T]$. 
\subsection{A general Longstaff-Schwartz algorithm}\label{sec:LongstaffSchwartz}  Replacing the continuous time interval $[0,T]$ by some finite grid $0=t_0<t_1<\dots < t_N=T$, the \emph{dynammic programming principle} \cite[Page eq. (2.1.7)]{peskir2006optimal} for the discrete optimal stopping reads \begin{equation}\label{eq:dynammicprorgamming}
Y_{t_N}=Z_{t_N}, \quad Y_{t_n} = \max ( Z_{t_n},\mathbb{E}[Y_{t_{n+1}}|\mathcal{F}^X_{t_n}] ), \quad 0 \leq n \leq N-1,
\end{equation} and $\tau^\star = \inf\{t_n: Z_{t_n} \geq \mathbb{E}[Y_{t_{n+1}}|\mathcal{F}^X_{t_n}] \}$ is optimal. Motivated by the famous Longstaff and Schwartz algorithm \cite{longstaff2001valuing}, the first optimal stopping time can be obtained as $\tau_0$ of the following recursion \begin{equation}\label{eq:LongstaffSchwartz}
\tau_N = t_N, \quad \tau_n = t_n1_{\{Z_{t_n} \geq \mathbb{E}[Z_{\tau_{n+1}}|\mathcal{F}^{X}_{t_n}]\}}+\tau_{n+1}1_{\{Z_{t_n} <  \mathbb{E}[Z_{\tau_{n+1}}|\mathcal{F}^{X}_{t_n}]\}}, \quad 0 \leq n \leq N-1.
\end{equation} The remaining question is, how to compute the \emph{continuation values} $\mathbb{E}[Z_{\tau_{n+1}}|\mathcal{F}^X_{t_n}]$ for possibly non-Markovian state-processes $X$, which will be the main topic of the forthcoming Section \ref{sec:SignatureStopping}. For now, assume we are given a suitable family of basis-functions $(\psi_k)$, such that $\mathbb{E}[Z_{\tau_{n+1}}|\mathcal{F}^X_{t_n}]\approx \sum_k \alpha_k^n\psi_k(X|_{[0,t_n]})$ with some coefficients $(\alpha^n_k)$.  A general version of the Longstaff-Schwartz algorithm goes as follows: \begin{enumerate}
\item Draw $i=1,\dots,M$ samples $X^{(i)}$ and $Z^{(i)}$ and initialize $\tau_N^{(i)} \equiv t_N$.
\item Then, for $n=N-1,\dots,1$ \begin{itemize}
\item solve the minimization problem $$\alpha^{\star,n}:= \argmin_{\alpha}\frac{1}{M}\sum_{i=1}^M \left (Z^{(i)}_{\tau_{n+1}^{(i)}}-\sum_k \alpha_k \psi_k^n(X|_{[0,t_n]}^{(i)}) \right )^2$$
\item Set $\tau_n^{(i)}=t_n$ if $Z^{(i)}_{t_n} \geq \sum_k \alpha^{\star,n}_k \psi_k^n(X|_{[0,t_n]}^{(i)})$, and $\tau_n^{(i)}=\tau_{n+1}^{(i)}$ otherwise.
\end{itemize} 
\item Finally, draw $i=1,\dots,M'$ independent sample-paths $\tilde{X}^{(i)}$ and $\tilde{Z}^{(i)}$, and compute the stopping times $\tilde{\tau}^{(i)}= \inf\left \{t_n: \tilde{Z}^{(i)}_{t_n} \geq \sum_k \alpha^{\star,n}_k \psi_k^n(\tilde{X}|_{[0,t_n]}^{(i)})\right \}$ and the lower-biased estimator $$\tilde{y}_0 = \max \left ( Z_{t_0},\frac{1}{M'}\sum_{i=1}^{M'} \tilde{Z}^{(i)}_{\tilde{\tau}^{(i)}} \right )$$
\end{enumerate} Notice that the resimulation in the final step ensures, that the $\tilde{\tau}^{(i)}$ are indeed stopping times, since the computed coefficients $\alpha$ are now independent of the samples, and thus $\max(Z_{t_0},\mathbb{E}[Z_{\tilde{\tau}_1}] )$ is a true lower-bound. The Monte-Carlo approximation $\tilde{y}_0$ is therefore lower-biased and strictly speaking not a true lower-bound, due to the Monte-Carlo error with respect to $M'$. Nevertheless, this error is typically chosen to be very small, and we will refer to $\tilde{y}_0$ as lower-bound anyways.

\subsection{A general dual procedure}\label{sec:dual_OSP}
While the Longstaff-Schwartz procedure described in the last section provides us with lower-bounds for the optimal value, it is desirable to additionally have an algorithm producing upper-bounds. To this end, it is useful to consider the (discretized) dual formulation of \eqref{eq:optimalstopping} due to \cite{rogers2002monte} \begin{equation}\label{eq:dual_optimalstopping}
\inf_{M\in \mathcal{M}^2}\mathbb{E}[\max_{0\leq k \leq N}(Z_{t_k}-M_{t_k})],
\end{equation} where the minimization is over discrete $L^2$-martingales with respect to the filtration $\mathcal{G}_n=\mathcal{F}^X_{t_n}, n=0,\dots,N$. By the Martingale Representation Theorem \cite[Theorem 4.5]{karatzas1991brownian}, if the underlying filtration $(\mathcal{F}^X_t)$ is generated by a Brownian motion $W$, such martingales are of the form $M_{t_k}^{\alpha}=\int_0^{t_k} \alpha_sdW_s$ for some $(\mathcal{F}_t^X)-$progressive process $\alpha$. On the other hand, due to the progressive measurability, the integrands are of the form $\alpha_t = f(X|_{[0,t]})$ for some (deterministic) function $f$. Assume again that we are given some family of basis-functions $(\psi_k)$ such that $f(X|_{[0,t]}) \approx \sum_k\beta_k\psi_k(X|_{[0,t]})$ for some coefficients $\beta$. If the family  $(\psi_k)$ is \emph{rich} enough, it is reasonable to parametrize the space $\mathcal{M}^2$ by the span of basis-martingales $\{M^k_t=\int_0^t\psi_k(X|_{[0,s]})dW_s: k \geq 0\}$, which leads to the following dual algorithm:
\begin{enumerate}
\item Draw $i=1,\dots,M$ sample-paths of $Z^{(i)},X^{(i)},W^{(i)}$, and compute the basis martingales $M^{k,(i)}$ (e.g. using an Euler-scheme)
\item Solve the minimization problem (see, e.g., \cite{desai2012pathwise,belomestny2013solving})$$\beta^{\star} = \argmin_\beta \frac{1}{M}\sum_{i=1}^M \max_{0\leq n \leq N}\left (Z^{(i)}_{t_n}-\sum_k \beta_k M^{k}_{t_n} \right ) $$
\item Finally, draw $i=1,\dots,M'$ independent sample-paths $\tilde{Z}^{(i)},\tilde{X}^{(i)},\tilde{W}^{(i)}$ and compute the martingales $\tilde{M}^{k,(i)}$ and the upper-biased estimator $$\tilde{y}_0= \frac{1}{M'}\sum_{i=1}^M \max_{0\leq n \leq N}\left (\tilde{Z}^{(i)}_{t_n}-\sum_k \beta^{\star}_k \tilde{M}^{k,(i)}_{t_n} \right )$$
\end{enumerate}
Similar as in the primal case, the independent resimulations ensure that we get true martingales $\tilde{M}^{\star}:=\sum_k \beta^{\star}_k\tilde{M}^k$, since $\beta^{\star}$ is independent of the samples, and therefore $\mathbb{E}[\max_k (Z_{t_k}-\tilde{M}^{\star}_{t_k})]$ is a true upper-bound. By the same abuse of terminology as in the primal case, we call the Monte-Carlo approximation $\tilde{y}_0$ upper-bound, keeping in mind that it is strictly speaking only upper-biased due to the Monte-Carlo error with respect to $M'$.

\section{Signature stopping policies and dual martingales}\label{sec:SignatureStopping}
To numerically solve the two algorithms presented in the last section, we require to learn functionals $f:\mathcal{X}\rightarrow \mathbb{R}$, where $\mathcal{X}$ is an infinite-dimensional path space. As motivated in the introduction, we will use the \emph{path-signature}, resp. the corresponding \emph{signature-kernel}, to tackle this problem. For the rough path theoretical details in this section, we refer to the excellent books \cite{friz2010multidimensional,friz2020course}, but see also \cite{cass2024lecture} with a focus on machine learning.

Let $(x(t):0 \leq t \leq T)$ be an $\mathbb{R}^d$-valued, $\alpha$-Hölder continuous path, $\alpha \in (0,1]$. The \emph{rough path signature} of $x$ is given by the path $\mathbbm{x}^{<\infty}_{\cdot, \cdot} : \Delta_{[0,T]} \to T((\R^d))$, defined (at least formally) as the collection of iterated integrals
\[
    \mathbbm{x}^{<\infty}_{s, t} := \sum_{n=0}^\infty \int_{s < t_1 < \cdots < t_n < t} d x(t_1) \otimes \cdots \otimes d x(t_n), \quad 0 \le s \le t \le T,
\]
taking values in the \emph{extended tensor algebra}
\[
    T((\R^d)) := \prod_{n=0}^{\infty} (\R^d)^{\otimes n},
\]
an algebra under a (naturally defined) non-commutative product $\otimes$. If $\alpha=1$, the meaning of the iterated integrals is in the sense of Riemann-Stieltjes, see \cite[Chapter 7.2]{friz2010multidimensional}, and less obvious becomes the case $\alpha<1$, since $d x$ does not have a meaning a-priori. However, Lyons' theory of rough paths \cite{lyons1998differential} allows us to give the latter a meaning, but higher-order information of the path is required. More precisely, we have to lift $x$ to a so-called $\mathbf{x}=(\mathbf{x}^{(1)},\dots,\mathbf{x}^{(N)}) \in \prod_{n=0}^{N} (\R^d)^{\otimes n}$, where $N= \lfloor\frac{1}{\alpha} \rfloor$ and $\mathbf{x}^{(1)}=x$, see \cite[Chapter 9]{friz2010multidimensional}. Having such a rough path lift $\mathbf{x}$ at hand, the notion of \emph{rough integration} allows to make sense of integrating against $d \mathbf{x}$, and a signature lift of $\mathbf{x}$ can be defined, sharing all the properties of the path-signature for smooth paths, which is due to Lyons' extension theorem \cite[Theorem 3.7]{lyons1998differential}.

\begin{remark}\label{rem:signature_assumption}
    The authors of \cite{boedihardjo2016signature} show that the map $x \mapsto \mathbbm{x}^{<\infty}$ is injective up to a certain equivalence class on path spaces, which can be eliminated when adding a monotone component to the path, e.g. the time-augmentation $\hat{x}(t)=(t,x(t))$. Moreover, by applying a so-called \emph{tensor-normalization} $\lambda :T((\R^d))\rightarrow T((\R^d))$ to the signature, the authors of \cite{chevyrev2018signature} introduce the notion of \emph{robust signatures} $\lambda(\mathbbm{x}^{<\infty})$, which can be seen as a bounded version of the signature, sharing all of its structural properties. For the theoretical results in this section, we always assume that our underlying process $X$ is already augmented by some monotone function, and this pair can be lifted to a \emph{geometric rough path} $\mathbf{X}$, see \cite[Definition 9.16]{friz2010multidimensional}, and we denote by $\mathbbm{X}^{<\infty}$ the unique robust rough path signature for some fixed normalization $\lambda$.
\end{remark}

\subsection{Linear signature learning}  
One of the most important algebraic properties of the signature is the following: For any two linear functionals on the tensor-algbera, that is $\ell_1,\ell_2 \in T((\mathbb{R}^d)^{\star})$, we can find $\ell_3$ (in fact, a closed-form expression can be given), such that $\ip{\ell_1}{\mathbbm{X}^{<\infty}_{0,T}} \ip{\ell_2}{\mathbbm{X}^{<\infty}_{0,T}} = \ip{\ell_3}{\mathbbm{X}^{<\infty}_{0,T}}$, for all signatures $\mathbbm{X}^{<\infty}_{0,T}$. In words, the space of linear functionals of the signature forms an algbera, which is point-seperating (since we assume to have a unique signature, see Remark \ref{rem:signature_assumption}). An application of a general Stone-Weierstrass result, see \cite[Section 2.1]{chevyrev2018signature}, tells us that the set of linear functional of the robust signature is dense in the set of bounded continuous functions on the corresponding path space. This fact can be used to deduce a global $L^p$-approximation for linear functionals of the signature, see \cite[Theorem 2.8]{bayer2023primal}, which then yields the following convergence results. We will always assume to be in the framework described in the beginning of Section \cite[Section 3.1]{bayer2023primal}, where certain assumptions on the path space, as well as on the payoff process are made precise. The following proposition summarizes the results \cite[Proposition 3.3 and 3.8]{bayer2023primal}, to which we refer for more precise statements and detailed proofs.\begin{proposition}\label{prop:linear}
Using the same notation from Section \ref{sec:Optimal stopping}, we consider $i=1,\dots,M$ independent sample-paths of $X,Z,W$, and we define the martingales $M^\ell_t= \int_0^t \ip{\ell}{\mathbbm{X}^{\leq K}_{0,s}}dW_s$ for some truncation level $K\in \mathbb{N}$. We define the estimators  \begin{equation}\label{eq:linearestimators}
\begin{aligned}
    \ell_P^{\star,n} & = \argmin_{\ell,|\ell| \leq K}\frac{1}{M}\sum_{i=1}^M \left (Z^{(i)}_{\tau_{n+1}^{(i)}}- \ip{\ell}{\mathbbm{X}^{\leq K,(i)}_{0,t_n}} \right )^2, \\  \ell_D^{\star} & = \argmin_{\ell,|\ell| \leq K} \frac{1}{M}\sum_{i=1}^M \max_{0\leq n \leq N}\left (Z^{(i)}_{t_n}-M^{\ell,(i)}_{t_n} \right ).
\end{aligned}
\end{equation} Then, for new samples $\tilde{X},\tilde{Z},\tilde{W}$, independent of $X,Z,W$, define the $(\mathcal{F}_t^X)-$stopping times $\tilde{\tau}^{(i)}= \inf\left \{t_n: \tilde{Z}^{(i)}_{t_n} \geq \ip{\ell_P^{\star,n}}{\mathbbm{X}^{\leq K,(i)}_{0,t_n}} \right \}$ and martingales $\tilde{M}^{\ell^\star_D,(i)}$. Then  \begin{equation}
\max \left ( \tilde{Z}_{t_0},\frac{1}{M}\sum_{i=1}^{M} \tilde{Z}^{(i)}_{\tilde{\tau}^{(i)}} \right ) \nearrow Y_0, \quad \frac{1}{M}\sum_{i=1}^M \max_{0\leq n \leq N}\left (\tilde{Z}^{(i)}_{t_n}-\tilde{M}^{\ell_D^{\star},(i)}_{t_n} \right ) \searrow Y_0,
\end{equation} as $M,K \rightarrow \infty$, where the convergence with respect to $M$ is almost sure convergence.
\end{proposition} 
While the last result ensures convergence for the primal and dual procedures, it does not come with quantitative statements about convergence rates. Additionally, these algorithms can become computationally expensive, as the size of the signature grows exponentially with $K$. In \cite{bayer2023primal} several numerical experiments were performed for these two algorithms in non-Markovian frameworks, in particular for models driven by fractional Brownian motion with small Hurst parameters. The accuracy of the method can be measured by the duality gap between lower and upper bounds, and in some examples these gaps were observed to be quite significant even for high truncation levels. Two important sources for this gap can be described as follows: First, especially in very rough regimes, the considered signature levels might not suffice to capture the relevant past of the non-Markovian processes, so that information is lost when truncating the signature. Secondly, the functionals on the path spaces we try to learn are highly non-linear, so that more sophisticated learning technologies are required to improve the performance. The goal of the next sections is to introduce non-linear extensions of the primal and dual algorithm, based on deep-, resp. kernel-learning methodologies, with the objective of improving the duality gap.
    
\subsection{Deep signature learning}\label{sec:DNNsignature}
Applying \emph{Deep Neural Networks} (DNNs) on top of the signature was considered several times in the literature, see for instance \cite{kidger2019deep}, where even more generally, the signature can act as a layer in a network. Here, we simply consider the truncated signature as the input for a DNN of the form \[\theta = \beta \circ \sigma \circ A_1 \circ \sigma \cdots \circ A_I,
\] where $\beta,A_0,\dots,A_I$ are affine, $\beta: \R^{m}\rightarrow \R, A_I: T((\mathbb{R}^d)) \rightarrow \R^{m}, A_{I-i}:\R^{m} \to \R^{m}$, for $i=1,\dots, I-1$, and $\sigma$ is an activation function. We call $\theta$ a \emph{signature DNN} with $I$ hidden layers, each of which having $m$ neurons. The signature map $\mathbf{x}|_{[0,t]}\mapsto \mathbbm{x}^{<\infty}_{0,t}$ represents the input layer, and the first hidden layer acting on the signature, can be seen as a vector of linear functionals $A_I = (\ell_1,\dots,\ell_m)$, and $A_I(\mathbbm{x}^{<\infty}) = (\ip{\ell_1}{\mathbbm{x}^{<\infty}},\dots,\ip{\ell_m}{\mathbbm{x}^{<\infty}})^\top \in \mathbb{R}^m$. The remaining hidden layers are all of the form $A_j(x) = A_jx+b_j$, where $A_j \in \mathbb{R}^{m \times m},b_j \in \mathbb{R}^m$, and the output layer is given by $\beta(x)=\beta_0+\sum_{j=1}^m\beta_jx_j$ where $\beta \in \mathbb{R}^{m+1}$. We denote by $\mathrm{DNN}^{\sigma,I}_{\mathrm{sig}}$ the set of all such signature DNNs $\theta$ with $I$ hidden layers, and activation function $\sigma$.

It is well known, that already $I=1$ hidden layer DNNs are universal, see, e.g. \cite{leshno1993multilayer}. For this reason, and to ease the notation for the theoretical results, in the remainder of this section we fix $I=1$. Nevertheless, all the results trivially extend to multiple hidden layers $I>1$, and in the numerical experiments we mostly choose more than one hidden layer. Setting $\mathrm{DNN}^{\sigma}_{\mathrm{sig}}:= \mathrm{DNN}^{\sigma,1}_{\mathrm{sig}}$, we can explicitly write the set as \begin{equation}\label{eq:DNN_set}
    \mathrm{DNN}^{\sigma}_{\mathrm{sig}} = \left \{ \mathbf{x}|_{[0,t]}\mapsto \beta_0+ \sum_{j=1}^m\beta_j \sigma \Big (\ip{\ell_j}{\mathbbm{x}^{<\infty}_{0,t}} \Big ): (\beta,\ell) \in \mathbb{R}^{m+1}\times (\mathcal{W}^{\leq k})^m, m,k \geq 1\right \},
\end{equation} where $\mathcal{W}^{\leq k}$ denotes the set of linear functionals, which only act on the truncated signature $\mathbbm{x}^{\leq k}$,  and $\mathbf{x}|_{[0,t]}$ are \emph{stopped $\alpha$-Hölder rough paths}, see  \cite[Section 2]{bayer2023primal} for details. For fixed $(m,k)\in \mathbb{N}^2$ and any pair $(\beta,\ell) \in \mathbb{R}^{m+1}\times (\mathcal{W}^{\leq k})^m$, denote by $\theta^{(\beta,\ell)}$ the corresponding functional in the set $\mathrm{DNN}^{\sigma}_{\mathrm{sig}}$. {\color{black} Moreover, for any sequence of real numbers $\eta = (\eta_m)_{m\in \mathbb{N}}$, let us define the following subsets \[
\mathcal{D}_\eta^{(m,k)} = \left \{\theta^{(\beta,\ell)}\in \mathrm{DNN}^{\sigma}_{\mathrm{sig}}:(\beta,\ell) \in \mathbb{R}^{m+1}\times (\mathcal{W}^{\leq k})^m, \Vert \theta^{(\beta,\ell)}\Vert  \leq \eta_m\right \},
\] where  $\Vert \theta^{(\beta,\ell)}\Vert:= \Vert \beta \Vert_{\mathbb{R}^{m+1}}+\Vert \ell \Vert_{T^{\leq k}(\mathbb{R}^d)}$. Whenever $\eta_m \to \infty$ as $m\to \infty$, which we will always assume, we can observe that \[\mathrm{DNN}^{\sigma}_{\mathrm{sig}} = \bigcup_{m,k \geq 1}\mathcal{D}_{\eta}^{(m,k)} \]  }Revisiting the Longstaff \& Schwartz algorithm presented in Section \ref{sec:LongstaffSchwartz}, it is tempting to define a sequence of \emph{deep stopping times} $(\tau^{(m,k)}_n: 1 \leq n \leq N)$, recursively defined similar to \eqref{eq:LongstaffSchwartz}, but with conditional expectations replaced by {\color{black} $\theta_P^{n,\star}(\mathbf{X}|_{[0,t_n]})$, where for $0\leq n \leq N$ \begin{equation}\label{eq:deep_primal_minimizer}
        \theta^{n,\star}_P  = \argmin_{\theta \in \mathcal{D}_\eta^{(m,k)}}\mathbb{E} \left [\big (Z_{\tau^{(m,k)}_{n+1}}- \theta(\mathbf{X}|_{[0,t_n]})\big)^2\right ].
    \end{equation}}
In words, we recursively learn the continuation values as neural networks of truncated signatures, {\color{black} where the weights are bounded by $\eta_m$, which ensures existence of the minimizer \eqref{eq:deep_primal_minimizer}.} Similarly, we approximate the Doob-martingale from Section \ref{sec:dual_OSP} by \emph{deep martingales} {\color{black} $M^{(m,k)}_t= \int_0^t\theta_D^{\star}(\mathbf{X}|_{[0,s]})dW_s$ , where \begin{equation}\label{eq:deep_dual_minimizer}
    \theta_D^\star  = \argmin_{\theta \in \mathcal{D}_\eta^{(m,k)}}\mathbb{E} \left [\max_{0\leq n \leq N}\bigg (Z_{t_n}-\int_0^{t_n}\theta(\mathbf{X}|_{[0,s]})dW_s \bigg) \right ].
    \end{equation}}
Thanks to the universality of both signatures and DNNs, the following result should not come as a surprise. Similar as in the last section, we adopt the assumptions described in the beginning of \cite[Section 3.1]{bayer2023primal}.
\begin{proposition}\label{prop:deep}
    Assuming that the filtration is Brownian, that is  $\mathcal{F}^X=\mathcal{F}^W$, and the activation function $\sigma$ is non-polynomial, we have \[ Y_0^N=\lim_{k,m \to \infty} \max \left (Z_{t_0},\mathbb{E}[Z_{\tau_1^{(m,k)}}] \right ) = \lim_{k,m \to \infty} \mathbb{E}\Big [\max_{0 \leq n \leq N}(Z_{t_n}-M^{(m,k)}_{t_n})\Big ].  \]  
\end{proposition}
An outline of the proof can be found in Appendix \ref{sec:appendix}. Similar to the linear case, in practice we solve the sample average approximation of \eqref{eq:deep_primal_minimizer}-\eqref{eq:deep_dual_minimizer}, that is{\color{black} \begin{equation}\label{eq:non-linearestimators}
\begin{aligned}
\theta_P^{n,\star} &= \argmin_{\theta \in \mathcal{D}_\eta^{(m,k)}}\frac{1}{M}\sum_{i=1}^M \left (Z^{(i)}_{\tau_{n+1}^{(i)}}- \theta(\mathbf{X}^{(i)}|_{[0,t_n]}) \right )^2, \\  \theta^\star_D & = \argmin_{\theta \in \mathcal{D}_\eta^{(m,k)}} \frac{1}{M}\sum_{i=1}^M \max_{0\leq n \leq N}\left (Z^{(i)}_{t_n}-M^{\theta,(i)}_{t_n} \right ),
\end{aligned}
\end{equation}} for i.i.d sample-paths $Z^{(i)},\mathbf{X}^{(i)},W^{(i)}$. The latter (non-convex) optimization problems can then be solved using stochastic gradient descent. More details about this and the DNN architecture will be discussed in Section \ref{sec:numerical_examples}. Let us conclude this section with a remark.
\begin{remark}\label{rmk:log-signature}
    To address the complexity issue coming from the size of the signature, it can be helpful to consider the so-called log-signature. The latter is defined by $\mathbb{L}^{<\infty}_{s,t}:= \mathrm{log}^{\otimes}(\mathbbm{X}^{<\infty}_{s,t})$, where $\mathrm{log}^\otimes$ is the bijection \[ \mathrm{log}^{\otimes}(1+\mathbf{x}) := \sum_{k \geq 0}\frac{(-1)^{k+1}}{k}\mathbf{x}^{\otimes k} \in T((\R^d)), \quad \mathbf{x}\in T((\R^d)), \mathbf{x}^{(0)}=0.
\] 
It can be shown that the dimension of the truncated log-signature $\mathbb{L}^{\leq K}$, grows much slower than the one of $\mathbbm{X}^{\leq K}$, see, e.g., \cite{reizenstein2017calculation}, and denoting by $\exp^{\otimes}$ the inverse of $\mathrm{log}^{\otimes}$, we have $\mathbbm{X}^{\leq K} = \exp^{\otimes}(\mathbb{L}^{\leq K})$. Although the log-signature itself is not longer universal, Proposition \ref{prop:deep} remains through for log-signature DNNs, that is replacing the signature in \eqref{eq:DNN_set} by its log-signature transform. In Remark \ref{rmk:log_sig_rmk}, after the proof of Proposition \ref{prop:deep}, we briefly explain why this is true and how to modify the proof.

\end{remark}


\subsection{Signature-kernel learning}\label{sec:signaturekernel}
Let us start by recalling some notions from general RKHS theory, see, e.g. \cite{steinwart2008support}. Given a feature map $\phi: \mathcal{X}\rightarrow F$, where $F$ is a Hilbert space (the so-called \emph{feature space}), one can define the associated kernel \begin{equation}\label{eq:general_kernel}
    k: \mathcal{X} \times \mathcal{X} \rightarrow \mathbb{R}, \quad k(x,y) := \ip{\phi(x)}{\phi(y)}_{F}.
\end{equation} If the kernel is positive-definite, there exists a unique RKHS $\mathcal{H}$ with reproducing kernel $k$, see \cite[Theorem 4.21]{steinwart2008support}, in the sense that for any $f\in \mathcal{H}$, we have $f(x) = \ip{k(\cdot,x)}{f}_{\mathcal{H}}$, and in particular we have $\mathcal{H}= \overline{\mathrm{span}\{k(\cdot,x): x\in \mathcal{X} \}}$. In our case, the feature map $\phi$, maps a path $x\in \mathcal{X}$ to its path-signature $\mathbbm{x}^{<\infty}$, and therefore the signature kernel is naturally defined by
\[
    k_{s,t}(x,y) := \ip{\mathbbm{x}^{<\infty}_{0,s}}{\mathbbm{y}^{<\infty}_{0,t}}, \quad 0 \leq s,t \leq T
\]
where $\ip{\cdot}{\cdot}$ is the natural extension of the inner products $(\mathbb{R}^d)^{\otimes k}$ to the tensor algebra $T((R^d))$, see \cite[Definition 2.1.1]{cass2024lecture}. It has been shown in \cite{salvi2021signature}, that the signature-kernel solves a Goursat-type PDE with respect to the time variables. This \emph{kernel trick} allows us to evaluate the signature-kernel, representing the whole signature, by numerically solving a second-order PDE. An important extension, designed for rougher inputs, was discussed in the recent work \cite{lemercier2024high}.

An important observation is that universality of the signature feature map is equivalent to the universality of the signature-kernel, see for instance \cite[Section 6]{chevyrev2018signature}, but also \cite[Chapter 2.1]{cass2024lecture}. 
Returning to optimal stopping, this motivates us to seek for functionals in the primal and dual problems, solving the regularized minimizing problems on the RKHS \begin{equation}\label{eq:RKHS_minimization}
    f^{\lambda} = \argmin_{f\in \mathcal{H}} \mathcal{L}(f) + \lambda \Vert f \Vert^2_{\mathcal{H}}, \quad \lambda > 0,
\end{equation} where the loss $\mathcal{L}$ is either the primal or the dual loss function. More precisely, in the primal case, at each exercise date $n$, the loss is simply the mean-square error  $\mathcal{L}_n(f) = \Vert Z_{\tau_{n+1}}-f(X|_{[0,t_n]}) \Vert_2^2$. In the dual case, it is given by $\mathcal{L}(f)= \Vert \max_{1\leq k \leq N} (Z_{t_k}-M^f_{t_k}) \Vert_1$, where $M_t^{f} = \int_0^tf(X|_{[0,s]})dW_s$. Then, replacing the losses by their sampled version, the general representer theorem \cite[Theorem 1]{scholkopf2001generalized} reveals that the minimizers are of the form $f^{\alpha}(x) = \sum_{i=1}^M\alpha_i k(X^{(i)},x)$. Thus, let us define the minimizers \begin{equation}\label{eq:primal_kernel_estimator}
\alpha_P^{\star,n}  := \argmin_{\alpha \in \R^M}\frac{1}{M}\sum_{i=1}^M \left (Z^{(i)}_{\tau_{n+1}^{(i)}}- \sum_{j=1}^M \alpha_jk_{t_n,t_n}(X^{(i)},X^{(j)}) \right )^2 + \lambda \Vert f^{\alpha} \Vert^2_{\mathcal{H}}, \quad \lambda >0,
\end{equation} for all exercise date $n=1,\cdots,N-1$. Similarly, 
 \begin{align}\label{eq:dual_kernel_estimator}
\alpha_D^{\star} := \argmin_{\alpha \in \R^M} \frac{1}{M}\sum_{i=1}^M \max_{0\leq n \leq N}\left (Z^{(i)}_{t_n}-\sum_{j=1}^M\alpha_jM^{(i),(j)}_{t_n} \right ) + \lambda \Vert f^\alpha \Vert^2_{\mathcal{H}},\quad \lambda >0,
\end{align} for the kernel-martingales $M^{(i),(j)}_t = \left (\int_0^tk_{s,s}(X^{j},X)dW_s\right )^{(i)}.$ The algorithms in Section \ref{sec:Optimal stopping} can then easily be translated to the kernel-learning framework, by replacing the minimizers in the second step accordingly.\begin{remark}\label{rem:convergence_kernel}
    We expect that the convergence results in Proposition \ref{prop:linear} can be obtained analogously in this kernel-learning framework, when sending $M \to \infty$ and $\lambda \to 0$. For instance in the dual case, it follows by universality of the signature-kernel, that the closure of the span of the family of kernel-martingales $\mathcal{M}^K= \left \{\int k_{s,s}(x,X)dW_s: x \in \mathcal{X}\right \}$ corresponds to the space of $L^2(\mathcal{F}^W)$-martingales. For the sample average approximation and existence of minimizers, one can then argue similar as in \cite[Appendix A.2.]{bayer2023primal}, and use the representer theorem to conclude.
\end{remark}
\begin{remark}\label{rem:advantage_kernel}
    The obvious advantage of this method is that no truncation for the feature map is necessary, and therefore theoretically it does not suffer from a loss of information, see Remark \ref{rem:signature_assumption}. Modulo evaluation of the signature-kernel, the approximation error only depends on the regularization $\lambda$ and the number of samples $M$. Moreover, at least for the mean-square loss, it seems possible to theoretically study convergence rates of such algorithms. This, however, involves a more precise understanding of the signature RKHS $\mathcal{H}$ and integral operators therein, a problem that is outside the scope of this paper, but planned for future research.
\end{remark}
Having mentioned the theoretical advantages of the kernel-method, evaluating the signature-kernel, which means solving the Goursat-PDE \cite{salvi2021signature}, becomes the main difficulty in this procedure. Due to the recursive nature of the regression problems in the Longstaff and Schwartz algorithm in Section \ref{sec:LongstaffSchwartz}, typically large sample size are required to ensure stability. Moreover, as we are especially interested in paths of low regularity, a fine discretization grid is required to solve the Goursat PDE. Combining these observations with the fact, that the kernel-ridge-regressions involve computing and inverting the Gram-matrices $(k_{t_n,t_n}(X^{(i)},X^{(j)})_{i,j})$, it becomes clear that a reduction of the computational costs is required.

To this end, we propose the following approach, related to the so-called \emph{Nyström-method} for kernel-learning \cite{drineas2005nystrom,williams2000using}. We randomly select $L \ll M$ subsamples, denoting by $I_L$ the set of those indices, and in both the primal and dual optimization problems  we restrict the minimization to functions of the form $f(\cdot) =\sum_{i \in I^L}\beta_ik(x^{i},\cdot)$. It follows that we only need to compute the $(M \times L)$-matrices \[ \mathbf{K}_{t_n}=(k_{t_n,t_n}(X^j,X^i):i \in I_L, j=1,\dots,M), \quad n=1,\dots,N. \] 
For the primal example, that is, for the kernel ridge regression, we can additionally observe that the explicit solutions is given by \[ \alpha_P^{\star,n} =(\mathbf{K}_{t_n}^T \mathbf{K}_{t_n}+M \lambda \mathbf{R}_{t_n})^{-1}\mathbf{K}_{t_n}^T Z_{\tau_{n+1}} 
\] where $\mathbf{R}_{t_n}= (k_{t_n,t_n}(x^{i},x^{j}):i,j\in I_L)$. Further details about the implementation will be presented in the next section.

\section{Pricing American options under rough volatility}\label{sec:numerical_examples}
In this section we test the signature-based procedures for the problem of pricing American options in rough volatility models. In such models, the asset-price dynamics is given by
\begin{equation}\label{eq:PriceDynamics}
S_0=s_0, \quad dS_t = rS_tdt+S_tv_t \left (\rho dW_r+\sqrt{1-\rho^2}dB_t\right ), \quad 0 < t \leq T,
\end{equation} where $W$ and $B$ are two independent Brownian motions, the volatility $(v_t)_{t\in[0,T]}$ is $(\mathcal{F}^W_t)-$adapted and continuous, $\rho \in [-1,1]$ and $r>0$ the interest rate. We denote by $X$ the log-price, that is $X_t=\mathrm{log}(S_t)$. Compared to classical diffusion models, we are interested in volatility process $(v_t)$ driven by fractional Brownian motion, turning both the volatility and the price into non-Markovian processes. We will focus on the following two, arguably most popular examples of rough volatility specifications. 
\begin{example}{(Rough Bergomi \cite{bayer2016pricing})}\label{ex:rBergomi}
The rough Bergomi volatility is given by \begin{equation*}
    v_t=\xi_0\mathcal{E}\left (\eta \int_0^t(t-s)^{H-\frac{1}{2}}dW_s \right ),
\end{equation*} where $\mathcal{E}$ denotes the stochastic exponential. In all the numerical examples, we will choose $\eta = 1.9$ and $\xi_0= 0.09$.
\end{example}
\begin{example}{(Rough Heston \cite{el2019characteristic})}\label{ex:rHeston} The rough Heston volatility, resp. its variance $V=v^2$, is defined as (weak) solution to the Volterra-type CIR equation \begin{equation}
    V_t = V_0+ \int_0^t (t-s)^{H-\frac{1}{2}}\lambda(\theta-V_s)ds+\int_0^t(t-s)^{H-\frac{1}{2}}\nu \sqrt{V_s}dW_s.
\end{equation} In all the numerical examples, we will choose $V_0 = \theta = 0.02,\nu = \lambda= 0.3$.
\end{example}
The problem of pricing American (resp. Bermudan) options consists of solving the discrete optimal stopping problem  \begin{equation}\label{eq:Bermudda}
Y_0^{N}=\sup_{\tau \in \mathcal{S}^N_0}E[e^{-r\tau}\phi(S_\tau)],
\end{equation} where $\mathcal{S}^N_0$ is the set of $(\mathcal{F}^S_t)-$stopping times taking values in the set of possible exercise-dates $\{t_0,\dots,t_N\}$, and $\phi$ is the payoff function. We will focus on so-called \emph{put-options}, which corresponds to the payoff function $\phi(x) = (K-x)^{+}$ for a given strike price $K$.

\subsection{Implementation details} The code accompanying this section can be found in \url{https://github.com/lucapelizzari/Optimal_Stopping_with_signatures}. 
Before discussing numerical experiments, let us specify in detail how the procedures introduced in Section \ref{sec:SignatureStopping} are implemented. All the signatures are computed using the package \emph{iisignature}, see \cite{reizenstein2018iisignature}. For the signature kernel, we rely on the PDE solvers from the package \emph{sigkernel} to obtain the kernel as finite-difference solution to the Goursat PDE derived in \cite{salvi2021signature}. For the simulation of the rough Bergomi, we use \url{https://github.com/ryanmccrickerd/rough_bergomi} related to \cite{mccrickerd2018turbocharging}, and for the simulation of rough Heston \url{https://github.com/SimonBreneis/approximations_to_fractional_stochastic_volterra_equations} related to \cite{bayer2024efficient}.

\subsubsection*{Linear signature stopping}
This approach was already studied in detail in \cite{bayer2023primal}, and we adopt the choices made there. In particular, for the primal procedure we use the signature of $\hat{X}_t=(t,X_t)$, and for the dual procedure the signature of $\hat{Z}_t = (t,X_t,\phi(X_t))$, and in both we add Laguerre polynomials $L(X_t,v_t)$ to the set of basis functions. For more details about this choice we refer to \cite[Section 4]{bayer2023primal}. 

\subsubsection*{Deep signature stopping}
Here we make a slightly different choice for the basis compared to the linear approach, exploiting both the universality of the DNNs and the signature. First, it was observed for instance in \cite{bank2023rough}, that the price $X$ has a partial Markovian nature in $B$, and only depends on the past through the non-Markovian volatility process $(v_t)_{t\in [0,T]}$. Leaving rigorous arguments to \cite{bank2023rough}, one can for instance note that $\mathrm{Law}(X_{t+h}|\mathcal{F}_t) = \mathrm{Law}(X^{t,x}_{t+h}|\mathcal{F}^v_t)|_{x=X_t}$, so that conditional expectations $\mathbb{E}[\phi(X_{t+\Delta t})|\mathcal{F}_t] = f(X_t,(v_s)_{s\leq t})$ for some measurable function $f$. To capture the relevant memory of the dynamics, in the primal case we lift the variance-augmented process $\hat{v}_t=(\langle X \rangle_t,v_t)= (\int_0^tv_u^2du,v_t)$ the signature $\hat{\mathbb{V}}^{<\infty}$. In the dual case we simply lift the time-augmentation $\hat{v}_t=(t,v_t)$. The partial Markovianity motivates to simply apply DNNs on $\{X_t,\hat{\mathbb{V}}_{0,t}^{\leq K}\}$.

Let us now specify the DNN architecture: For the Longstaff and Schwartz algorithm, similar to \cite{lapeyre2021neural}, we rely on the \emph{Leaky ReLu} activation function $\sigma(x) = 1_{\{x\geq 0\}}x+1_{\{x<0\}}0.3x$ and the ADAM optimizer to fit the models at each exercise date, with a batch-size of $b=128$ and learning rate $\lambda = 10^{-3}$. Inspired by \cite{lapeyre2021neural}, we use $e=15$ epochs to learn the conditional expectation at the last exercise date (first regression in the algorithm), and then use the trained weights to initialize the DNN at the next exercise-date. Doing this allows to reduce the epochs to $e=1$ for $1\leq n < N-1$, which in turn reduces the computation time significantly. It is sufficient to consider $I=2$ hidden layers with each $D_i \equiv (\mathrm{dim}(\hat{\mathbb{V}}^{\leq K})+1)+32$ neurons. For the dual algorithm we rely on the classical \emph{ReLu} activation function $\sigma(x) = \max(x,0)$, and again using the ADAM optimizer with batch size $b=128$ and learning rate $10^{-3}$ to fit the model. Compared to the primal algorithm,  deeper neural networks with $I=6$ layers and $D_i \equiv (\mathrm{dim}(\hat{\mathbb{V}}^{\leq K})+1)+32$ are required, which is due to the non-linear nature of the integrand of the Doob-martingale.

\subsubsection*{Signature-kernel stopping}
For the kernel ridge regressions in the Longstaff and Schwartz algorithm based methods, we choose the kernel of the time-augmented path $\hat{N}_t=(\langle X \rangle_t,W_t,X_t)$. At each exercise date $n$, we randomly select $L=32$ subsamples, where the $i$-th sample is selected with probability $p_i = \frac{k_{t_n,t_n}(\hat{N}^{(i)},\hat{N}^{(i)})}{\sum_{j=1}^Mk_{t_n,t_n}(\hat{N}^{(j)},\hat{N}^{(j)})}$, noting that the latter only requires the diagonal of the Gram matrix $\mathbf{K}$. 

For the dual procedure, we randomly select the subsamples according to the probabilities, build on the quadratic variation of the kernel martingales $\langle \int_0^\cdot k_{s,s}(\hat{N}^{(i)},\hat{N})dW_s \rangle_T^{(i)} = \int_0^Tk_{s,s}(\hat{N}^{(i)},\hat{N}^{(i)})^2ds$, that is \[
p_i = \frac{\int_0^Tk_{s,s}(\hat{N}^{(i)},\hat{N}^{(i)})^2ds}{\sum_{j=1}^M\int_0^Tk_{s,s}(\hat{N}^{(j)},\hat{N}^{(j)})^2ds}.
\] Having $\mathbf{K}\in \mathbb{R}^{M \times L}$ at hand, we noticed that best performance can be achieved when solving \eqref{eq:dual_kernel_estimator} using a simple version of the neural network technology before, with $1$ hidden-layer and \emph{ReLu} activation.

\subsection{American put option prices in rough Bergomi and rough Heston}

In this section we compare all the signature-based methods to price Bermudan put options. In Table \ref{tab:putIntervals_08_bergomi}-\ref{tab:putIntervals_H07_Bergomi}, we present the rough Bergomi model with Hurst parameters $H=0.8$, resp. $H=0.07$, and for a range of strikes $K\in \{70,80,\dots,120\}$ and $S_0=100$. We choose the contract duration of $T=1$ year and $J=12$ early exercise options and interest rate $r=0.05$ and correlation $\rho = -0.9$. In the first column we report the European price, that is, the price for no early exercise opportunity, $\mathbf{E}= \mathbb{E}[e^{-rT}(K-S_T)^+]$. The second columns correspond to lower-bounds obtained in \cite{goudenege2020machine}, see also \cite{bayer2020pricing}. Finally, in the remaining columns, we compare the \emph{point estimate}, which is the biased-estimator from the Longstaff and Schwartz algorithm based on the training samples, with the lower- and upper-bounds obtained from independent testing samples. While the intervals in the linear case are taken from \cite{bayer2023primal}, we derive the intervals for the deep signature using truncation level $K=4$, discretization for the signature and martingales $N=600$, and both $M=12^{18}$ samples for training and testing, to ensure stability of the procedure. For the signature-kernel, we use $M=2^{17}$ samples, with $N=240$ discretization points for solving the Goursat PDE. For each strike, we highlight the best lower-, resp. upper-bound, and in the last column present the relative duality gap with respect to these two values. In Table \ref{tab:putIntervals_H01_Heston} we present the same considerations for the rough Heston model with $H=0.1,r=0.06$ and $\rho = -0.7$, a choice motivated in \cite{bayer2024efficient}, with smaller regularization parameter.  The trainings were repeated $20$ times and the overall Monte-Carlo error is below $0.01$.  Finally, in Table \ref{tab:comp_times} we compare the computational time (in seconds) of one training each, as well as the offline computation of the signature, resp. signature-kernel. \begin{figure}[h!]
    \centering
    \begin{subfigure}[t]{0.45\textwidth}
        \centering
        \includegraphics[width=\textwidth]{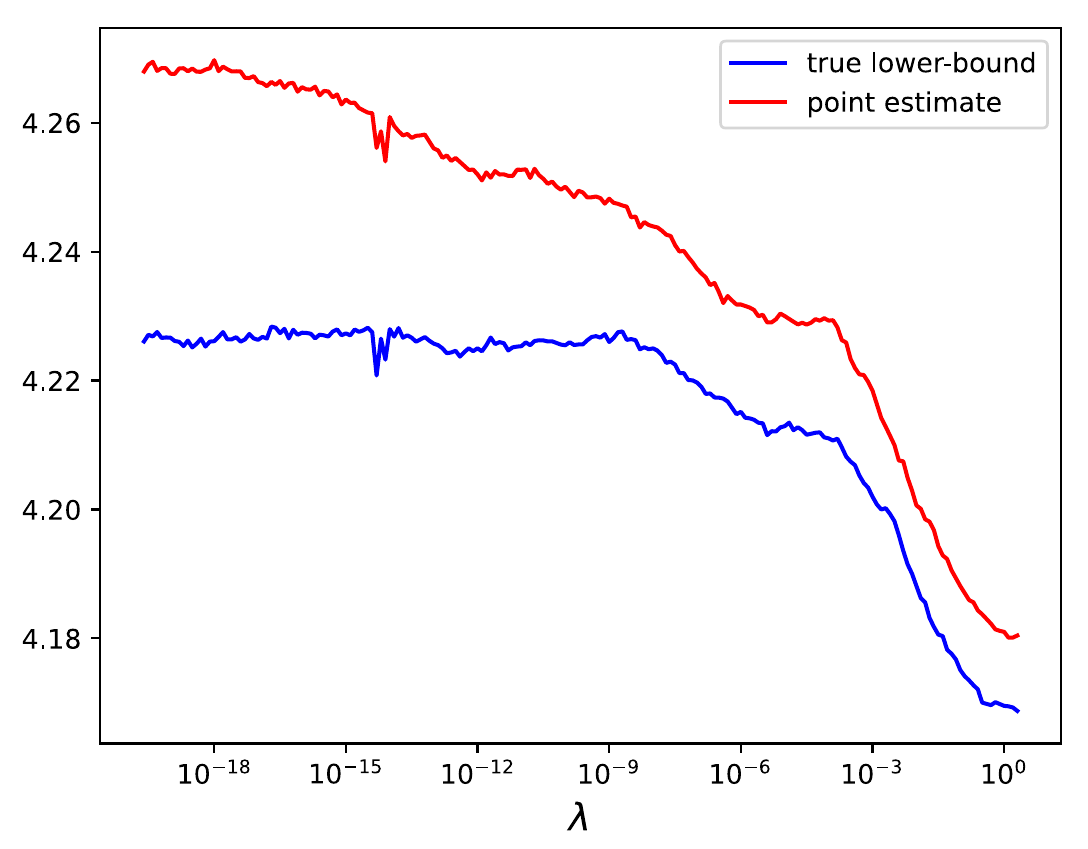}
        \caption{Rough Heston with $H=0.1$}
        \label{fig:plot1}
    \end{subfigure}%
    \hfill
    \begin{subfigure}[t]{0.45\textwidth}
        \centering
        \includegraphics[width=\textwidth]{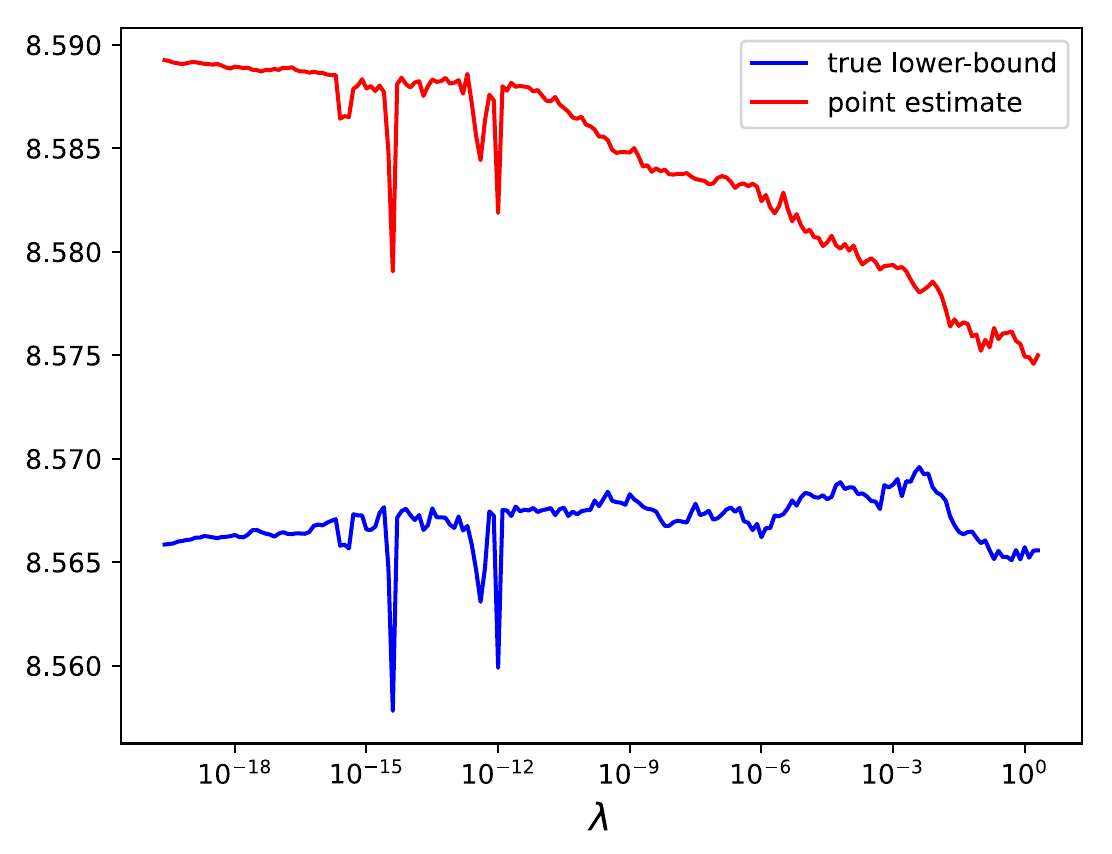}
        \caption{Rough Bergomi with $H=0.07$}
        \label{fig:plot2}
    \end{subfigure}
    \caption{Primal put option prices with strike $K=100$ and signature-kernel procedure, with respect to the penalization parameters $\lambda$. }
    \label{fig:ridge_plots}
\end{figure} In Figure \ref{fig:ridge_plots} we present the dependence of the lower-bounds, resp. point estimates with respect to $\lambda \in [0,2]$. The optimal choice is made for highest possible lower bounds (blue lines), so that the distance to the point estimate is reasonable, since larger difference suggest overfitting. According to this remark, we choose $\lambda = 10^{-3}$ for rough Bergomi, and $\lambda = 10^{-8}$ in the rough Heston model. 

First and most visibly, we observe that the deep signature method delivers significantly better upper-bounds than both other methods, with reasonable training and offline costs, see Table \ref{tab:comp_times}. Moreover, the deep method also mostly produces the smallest duality gap, that is the smallest distance between lower- and upper-bound. However, one can also observe that for both the point estimates and lower-bounds, the linear and kernel methods, which are based on conventional convex optimization problems, slightly outperform the deep signature method in general. It should be noted that, however, this comes at the price of offline computational costs, see Table \ref{tab:comp_times}, where we recall that the DNNs allow to only lift the augmented volatility process. As discussed in \cite{bayer2023primal}, for the linear procedure it is necessary to lift the three-dimensional path $(t,X_t,\phi(X_t))$ and add polynomials of the state, which explains the increased computational time in the linear case. While the signature-kernel method improves the linear approach, the complexity of deriving the Gram-matrix, i.e. solving the Goursat-PDE, see Table \ref{tab:comp_times}, leaves the signature-kernel procedure as the most expensive approach to solve the optimal stopping problem. To handle large data sets, it is therefore necessary to solve the complexity issue for the kernel, and   
a potential direction could be the \emph{random Fourier signature features} by \cite{toth2023random}, but this is outside the scope of this paper.

In summary, for our goal of achieving minimal duality gaps for arbitrary large sample sizes, the best possible results can be achieved by combining a Longstaff \& Schwartz algorithm based on the linear signature or the signature-kernel, exploiting the simplicity of the training procedure, together with the deep-signature approach for the upper-bounds, to capture the non-linearity of the Doob martingale integrand. If one has to choose one method, we recommend the deep-signature method, simply as it shows the smallest duality gaps and overall computational times.

\vspace{2mm}

\begin{table}[h!]
\centering
\resizebox{\textwidth}{!}{%
\ra{1.3}
\begin{tabular}{@{}ccccccccccccc@{}}
\toprule
\textbf{Strike} & \textbf{$\mathbf{E}$} & \textbf{\cite{goudenege2020machine}} &
\multicolumn{2}{c}{\textbf{Linear \cite{bayer2023primal}}} & 
\multicolumn{2}{c}{\textbf{Deep signature}} & 
\multicolumn{2}{c}{\textbf{Signature kernel}} & \textbf{Best Gap} \\ 
\cmidrule(lr){4-5} \cmidrule(lr){6-7} \cmidrule(lr){8-9}
& & & \textbf{Point est.} & \textbf{C.I.} & \textbf{Point est.} & \textbf{C.I.} & \textbf{Point est.} & \textbf{C.I.} & \\ \midrule
70 & 1.80 & 1.84 & 1.847 & [\textbf{1.83}, 1.90] & 1.831 & [1.82, \textbf{1.85}] & 1.835 & [\textbf{1.83}, 1.89] & 1.08 \% \\
80 & 3.02 & 3.10 & 3.101 & [\textbf{3.08}, 3.19] & 3.086 & [3.07, \textbf{3.12}] & 3.080 & [\textbf{3.08}, 3.16] & 1.28 \% \\
90 & 4.90 & 5.08 & 5.086 & [\textbf{5.07}, 5.17] & 5.058 & [5.04, \textbf{5.08}] & 5.061 & [5.06, 5.15] & 0.78 \% \\
100 & 7.73 & 8.19 & 8.188 & [\textbf{8.15}, 8.27] & 8.132 & [8.11, \textbf{8.17}] & 8.159 & [\textbf{8.15}, 8.26] & 0.24 \% \\
110 & 11.86 & 13.00 & 12.991 & [\textbf{12.97}, 13.09] & 12.922 & [12.87, \textbf{12.98}] & 12.944 & [12.94, 13.03] & 0.07 \% \\
120 & 17.57 & 20.28 & 20.219 & [\textbf{20.21}, 20.51] & 20.161 & [20.14, \textbf{20.25}] & 20.162 & [20.16, 20.27] & 0.19 \% \\ 
\bottomrule
\end{tabular}%
}
\caption{Comparison of Put option prices in rough Bergomi with {$H=0.8$}}
\label{tab:putIntervals_08_bergomi}
\end{table}

\begin{table}[h!]
\centering
\resizebox{\textwidth}{!}{%
\ra{1.3}
\begin{tabular}{@{}ccccccccccccc@{}}
\toprule
\textbf{Strike} & \textbf{E} & \textbf{\cite{goudenege2020machine}} &
\multicolumn{2}{c}{\textbf{Linear \cite{bayer2023primal}}} & 
\multicolumn{2}{c}{\textbf{Deep signature}} & 
\multicolumn{2}{c}{\textbf{Signature kernel}} & \textbf{Best Gap} \\ 
\cmidrule(lr){4-5} \cmidrule(lr){6-7} \cmidrule(lr){8-9}
& & & \textbf{Point est.} & \textbf{Interval} & \textbf{Point est.} & \textbf{Interval} & \textbf{Point est.} & \textbf{Interval} & \\ \midrule
70 & 1.83 & 1.88 & 1.929 & [\textbf{1.92}, 1.99] & 1.921 & [1.91, \textbf{1.95}] & 1.926 & [\textbf{1.92}, 2.01] & 1.53 \% \\
80 & 3.12 & 3.25 & 3.289 & [\textbf{3.27}, 3.37] & 3.281 & [3.26, \textbf{3.31}] & 3.286 & [\textbf{3.27}, 3.36] & 1.20 \% \\
90 & 5.08 & 5.34 & 5.394 & [\textbf{5.37}, 5.50] & 5.383 & [5.35, \textbf{5.44}] & 5.397 & [\textbf{5.37}, 5.53] & 1.28 \% \\
100 & 7.99 & 8.53 & 8.586 & [\textbf{8.57}, 8.77] & 8.555 & [8.52, \textbf{8.66}] & 8.589 & [8.56, 8.75] & 1.03 \% \\
110 & 12.17 & 13.28 & 13.314 & [\textbf{13.29}, 13.59] & 13.281 & [13.21, \textbf{13.45}] & 13.326 & [13.27, 13.46] & 1.18 \% \\
120 & 17.90 & 20.20 & 20.267 & [\textbf{20.24}, 20.66] & 20.163 & [20.14, 20.63] & 20.276 & [\textbf{20.24}, \textbf{20.44}] & 0.97 \%  \\ 
\bottomrule
\end{tabular}%
}
\caption{Comparison of Put option prices in rough Bergomi with {$H=0.07$}.}
\label{tab:putIntervals_H07_Bergomi}
\end{table}

\begin{table}[h!]
\centering
\resizebox{\textwidth}{!}{%
\ra{1.3}
\begin{tabular}{@{}cccccccccccc@{}}
\toprule
\textbf{Strike} & \textbf{$\mathbf{E}$} & 
\multicolumn{2}{c}{\textbf{Linear \cite{bayer2023primal}}} & 
\multicolumn{2}{c}{\textbf{Deep signature}} & 
\multicolumn{2}{c}{\textbf{Signature kernel}} & \textbf{Best Gap} \\ 
\cmidrule(lr){3-4} \cmidrule(lr){5-6} \cmidrule(lr){7-8}
& & \textbf{Point est.} & \textbf{Interval} & \textbf{Point est.} & \textbf{Interval} & \textbf{Point est.} & \textbf{Interval} & \\ \midrule
70 & 0.16 & 0.438 & [0.42, 0.53] & 0.426 & [0.42, \textbf{0.44}] & 0.430 & [\textbf{0.43}, 0.46] & 2.27 \% \\
80 & 0.42 & 0.966 & [0.94, 1.11] & 0.961 & [0.95, \textbf{0.97}] & 0.960 & [\textbf{0.96}, 1.01] &  1.03 \% \\
90 & 1.02 & 2.034 & [1.99, 2.28] & 2.031 & [\textbf{2.02}, \textbf{2.07}] & 2.027 & [\textbf{2.02}, 2.11] & 2.42 \% \\
100 & 2.66 & 4.245 & [4.16, 4.66] & 4.248 & [\textbf{4.24}, \textbf{4.36}] & 4.219 & [4.21, 4.44] & 2.75 \% \\
110 & 8.05 & 9.715 & [9.63, 10.52] & 9.660 & [9.66, 10.42] & 9.693 & [\textbf{9.68}, \textbf{10.14}] & 4.53 \% \\
120 & 16.68 & 19.500 & [19.49, 20.04] & 19.487 & [19.48, 20.00] & 19.502 & [\textbf{19.50}, \textbf{19.56}] & 0.30 \% \\ 
\bottomrule
\end{tabular}%
}
\caption{Comparison of Put option prices in rough Heston with {$H=0.1$}.}
\label{tab:putIntervals_H01_Heston}
\end{table}

\begin{table}[h!]
\centering
\ra{1.3}
\begin{tabular}{@{}cccc@{}}
\toprule
\textbf{Method} & \textbf{Training primal} &\textbf{Training dual} &  \textbf{Offline costs} \\ \midrule
Linear & 1.73 & 610.10 & 508.32   \\
Deep & 100.23 & 520.34 & 50.92  \\
Kernel &  1.69 & 390.98 &  5564.53\\
\bottomrule
\end{tabular}
\caption{Computational times (in seconds) for training and evaluation, and offline computation times (in seconds) for signatures and kernels }
\label{tab:comp_times}
\end{table}

\begin{figure}[h!]
    \centering
    \begin{subfigure}[t]{0.45\textwidth}
        \centering
        \includegraphics[width=\textwidth]{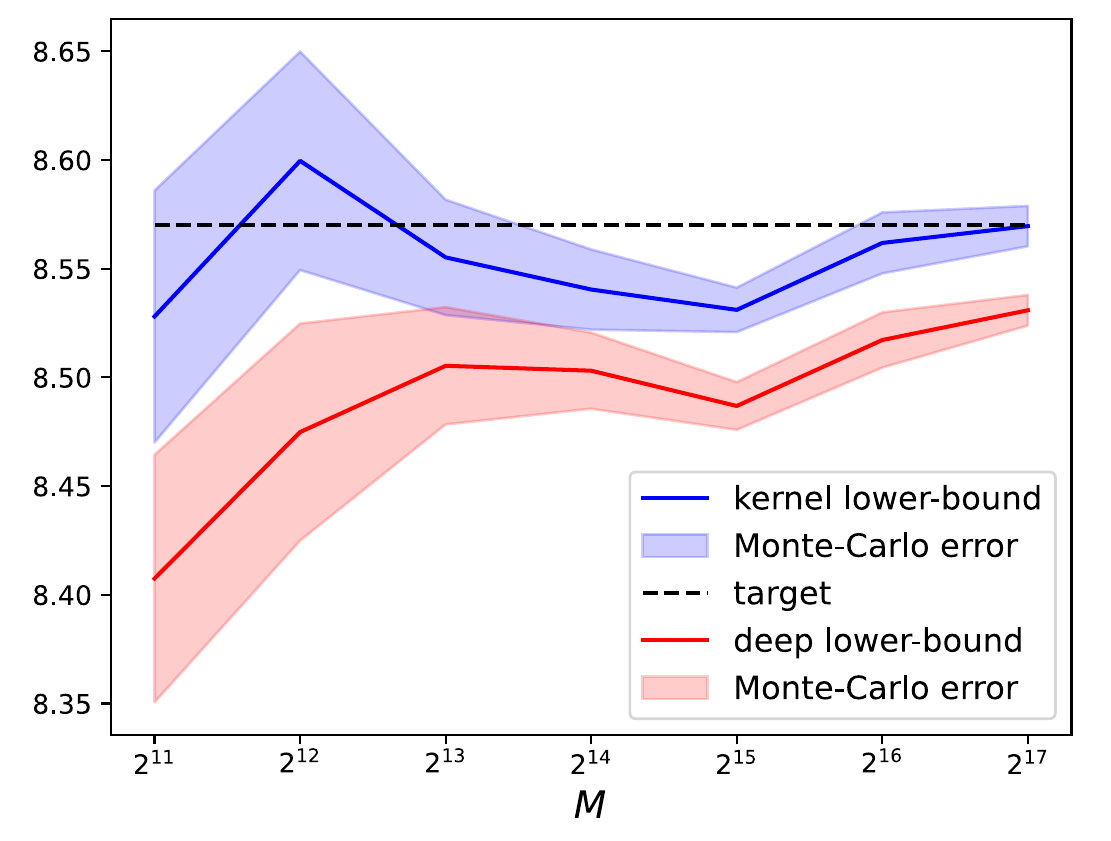}
        \caption{Primal}
    \end{subfigure}%
    \hfill
    \begin{subfigure}[t]{0.45\textwidth}
        \centering
        \includegraphics[width=\textwidth]{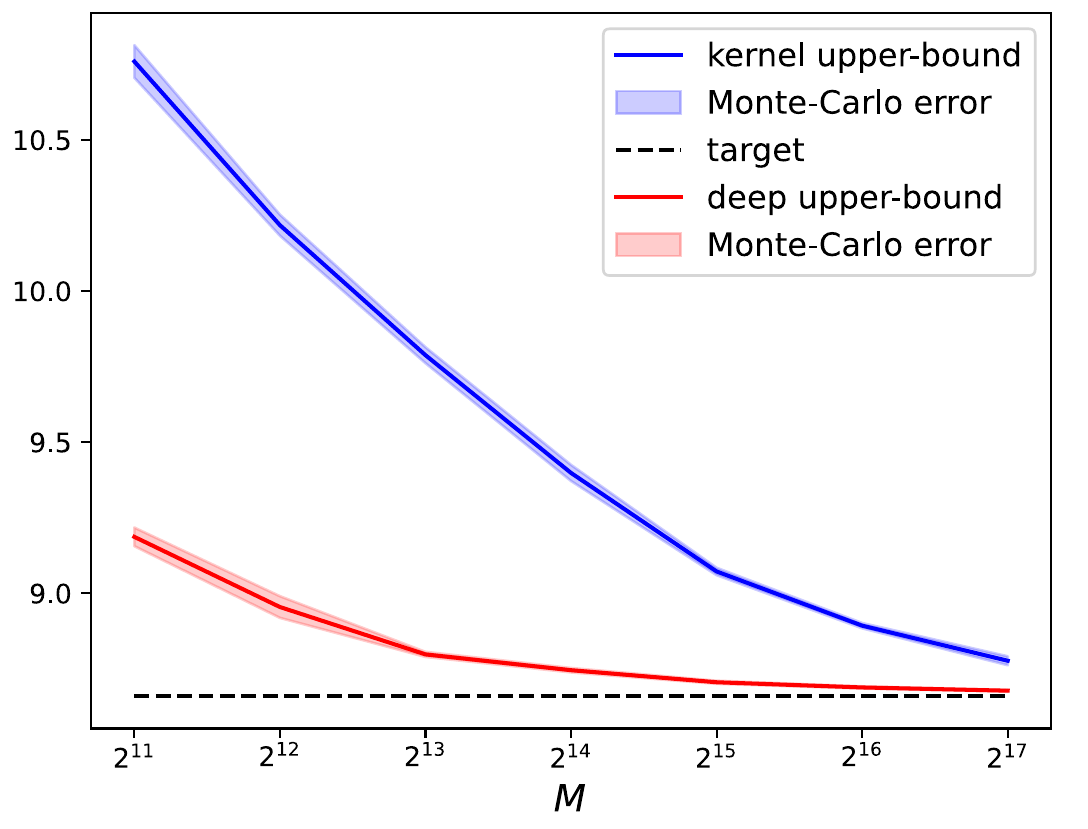}
        \caption{Dual}
    \end{subfigure}
    \caption{Put option prices for $H=0.07$ and strike $K=100$, with respect to training sample size $M$.}
    \label{fig:combined_sample_plots}
\end{figure}

On the other hand, it is important to note that we observe the primal signature-kernel method to be more stable compared to deep-signatures, with respect to smaller data sets. This unsurprising advantage of the kernel-method can become important when one works with real-word rather than synthetic data, where the number of samples cannot be arbitrary large, in which case the signature-kernel might be favorable. This is illustrated in Figure \ref{fig:combined_sample_plots} (a) for the rough Bergomi model with $H=0.07$, where we can observe more severe instability for the neural network training for smaller $M$, both in the sense of small lower-bounds and high training variance. The Monte-Carlo errors reflect this variance, which is obtained after independently repeating the training $20$ times for each sample size. For completeness, we show a similar plot in the dual case in Figure \ref{fig:combined_sample_plots} (b), where it is confirmed again that the deep-signature method heavily outperforms the signature-kernel approach, also in every data-size regime. The variance in the dual procedure is much smaller, as we only solve one optimization problem in the training, rather than at each exercise date.

\begin{figure}[h!]
    \centering
    \begin{subfigure}[t]{0.45\textwidth}
        \centering
        \includegraphics[width=\textwidth]{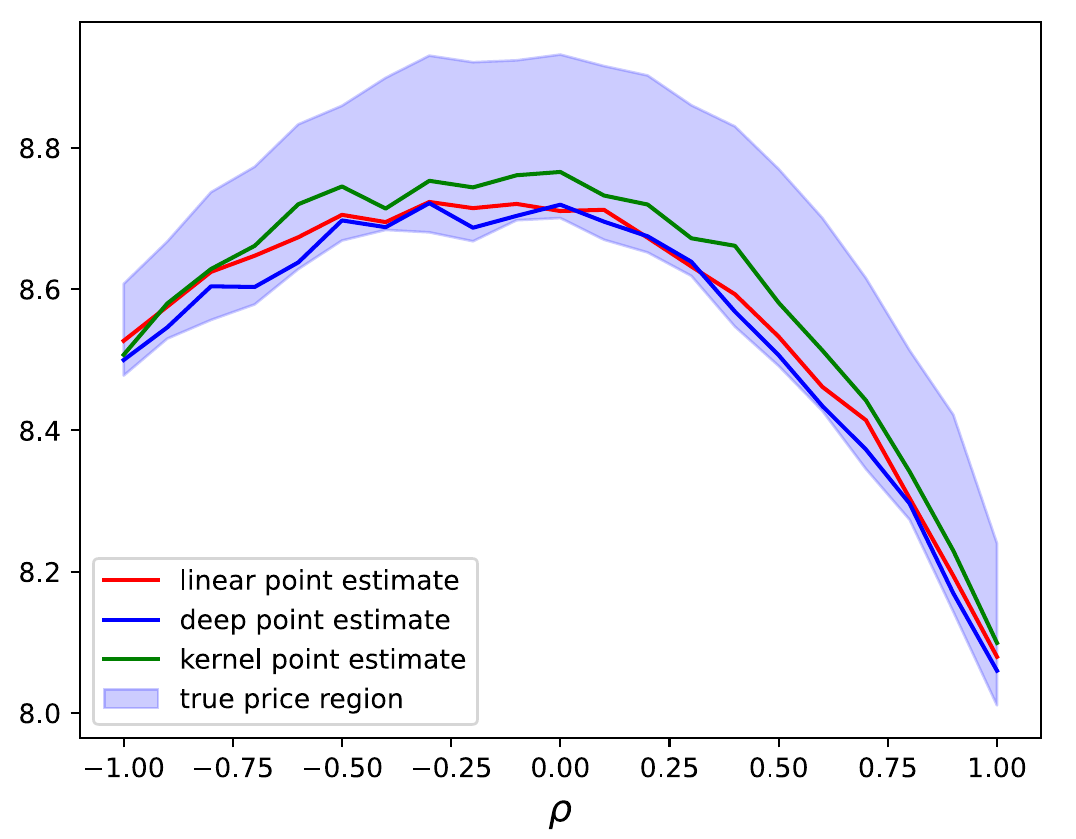}
        \caption{$H=0.07$}
        
    \end{subfigure}%
    \hfill
    \begin{subfigure}[t]{0.45\textwidth}
        \centering
        \includegraphics[width=\textwidth]{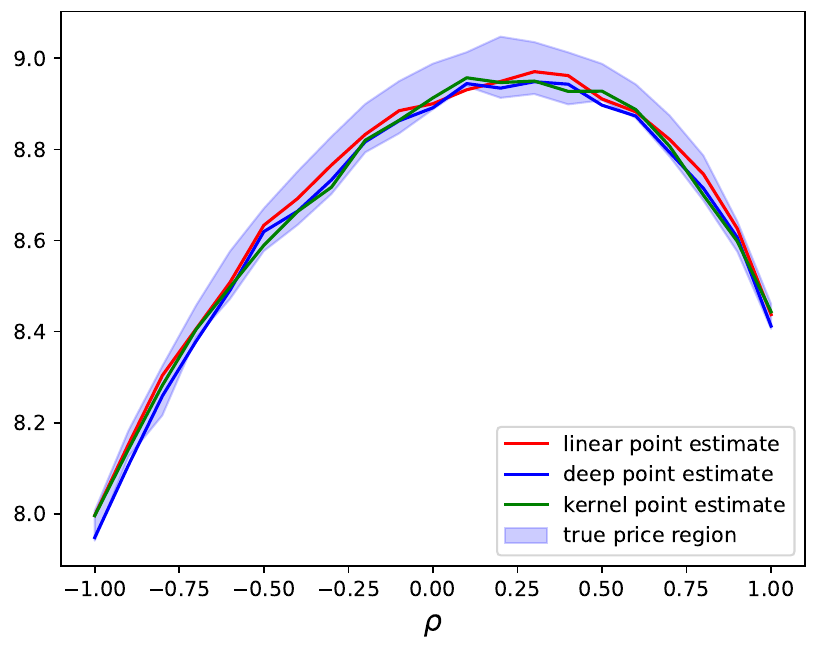}
        \caption{$H=0.8$}
        
    \end{subfigure}
    \caption{Point-estimates and pricing intervals for put option with $K=100$ in the rough Bergomi model, with respect to the correlation parameters $\rho \in [-1,1]$. }
    \label{fig:correlation_plots_bergomi}
\end{figure}

In Figure \ref{fig:correlation_plots_bergomi} we compare the options prices in the rough Bergomi model, with respect to the correlation parameter $\rho \in [-1,1]$, for the two (very different regimes) regimes $H=0.07$ and $H=0.8$. The blue region reflects the pricing interval derived from the deep-signature approach, and once again we present the point-estimates of all methods. In both cases we can observe, that the pricing method perform worse -- in the sense of duality gap -- when we come closer to the no-correlation case $\rho=0$, while the intervals get tighter in more correlated regimes. 

\begin{figure} 
    \centering
    \includegraphics[width=0.5\linewidth]{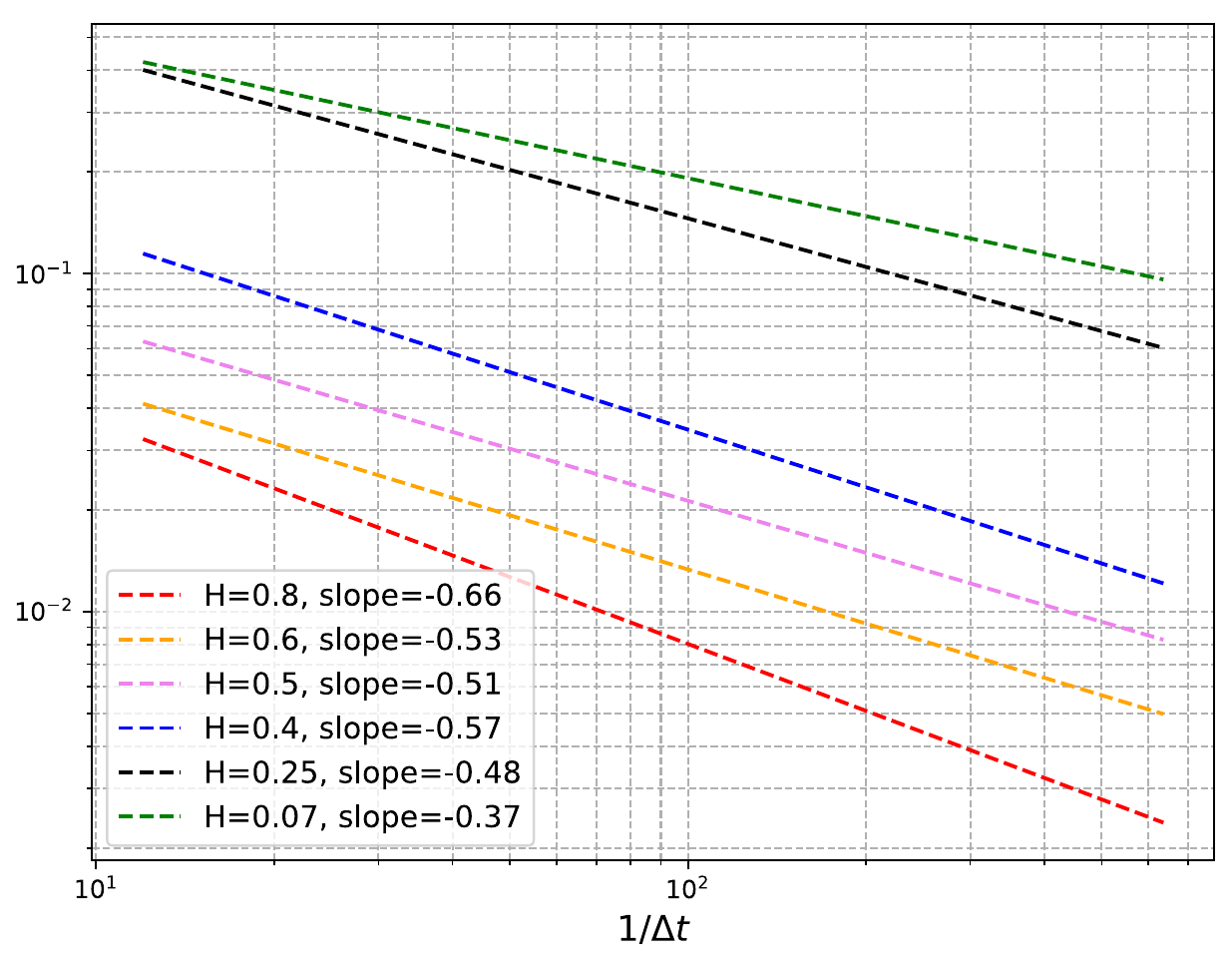}
    \caption{Linearly fitted duality gaps with respect to time-discretization and different Hurst regimes in the rough Bergomi model.}
    \label{fig:discretization}
\end{figure}

\begin{figure}[h!]
    \centering
    \begin{subfigure}[t]{0.45\textwidth}
        \centering
        \includegraphics[width=\textwidth]{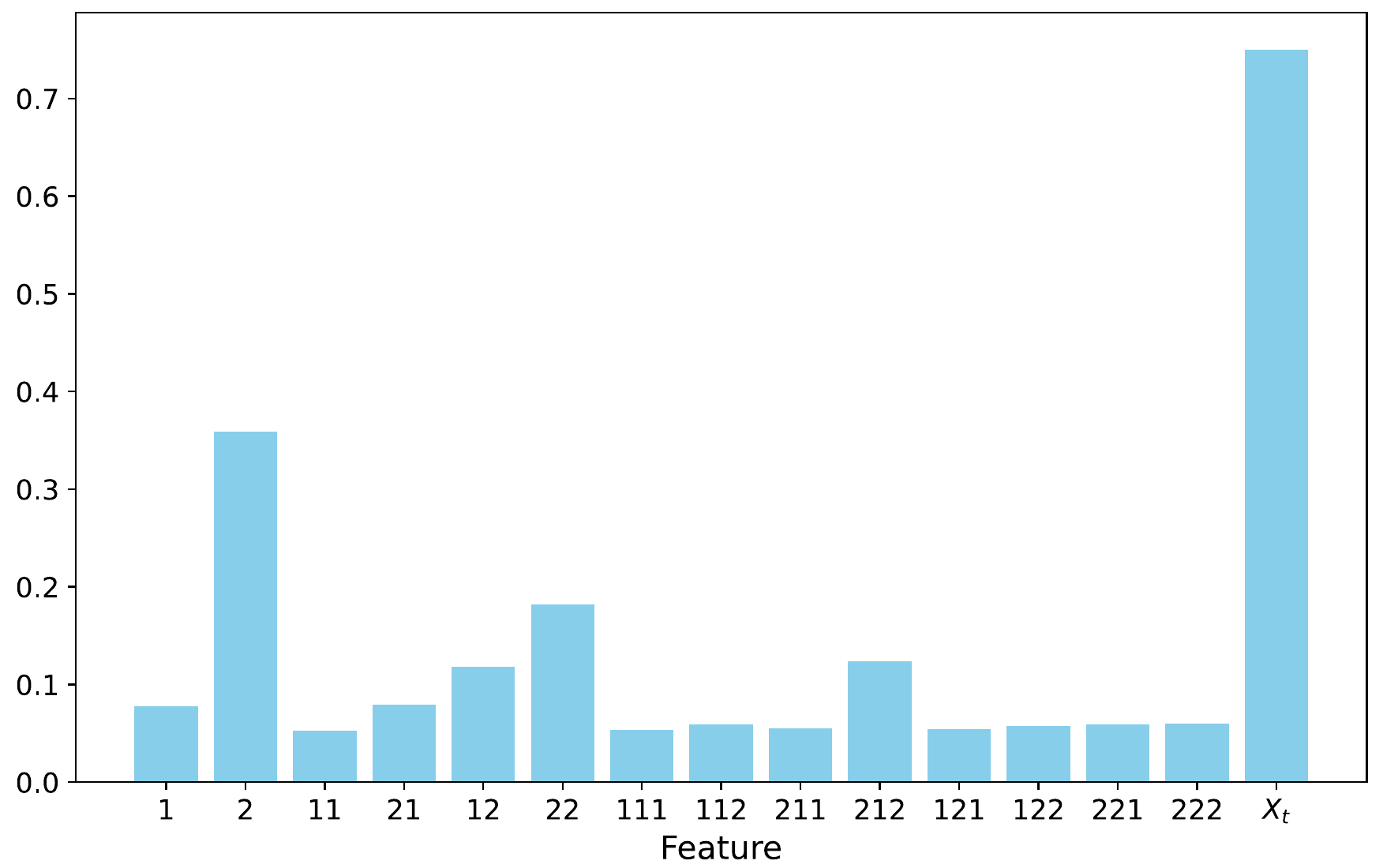}
        \caption{Primal}
        
    \end{subfigure}%
    \hfill
    \begin{subfigure}[t]{0.45\textwidth}
        \centering
        \includegraphics[width=\textwidth]{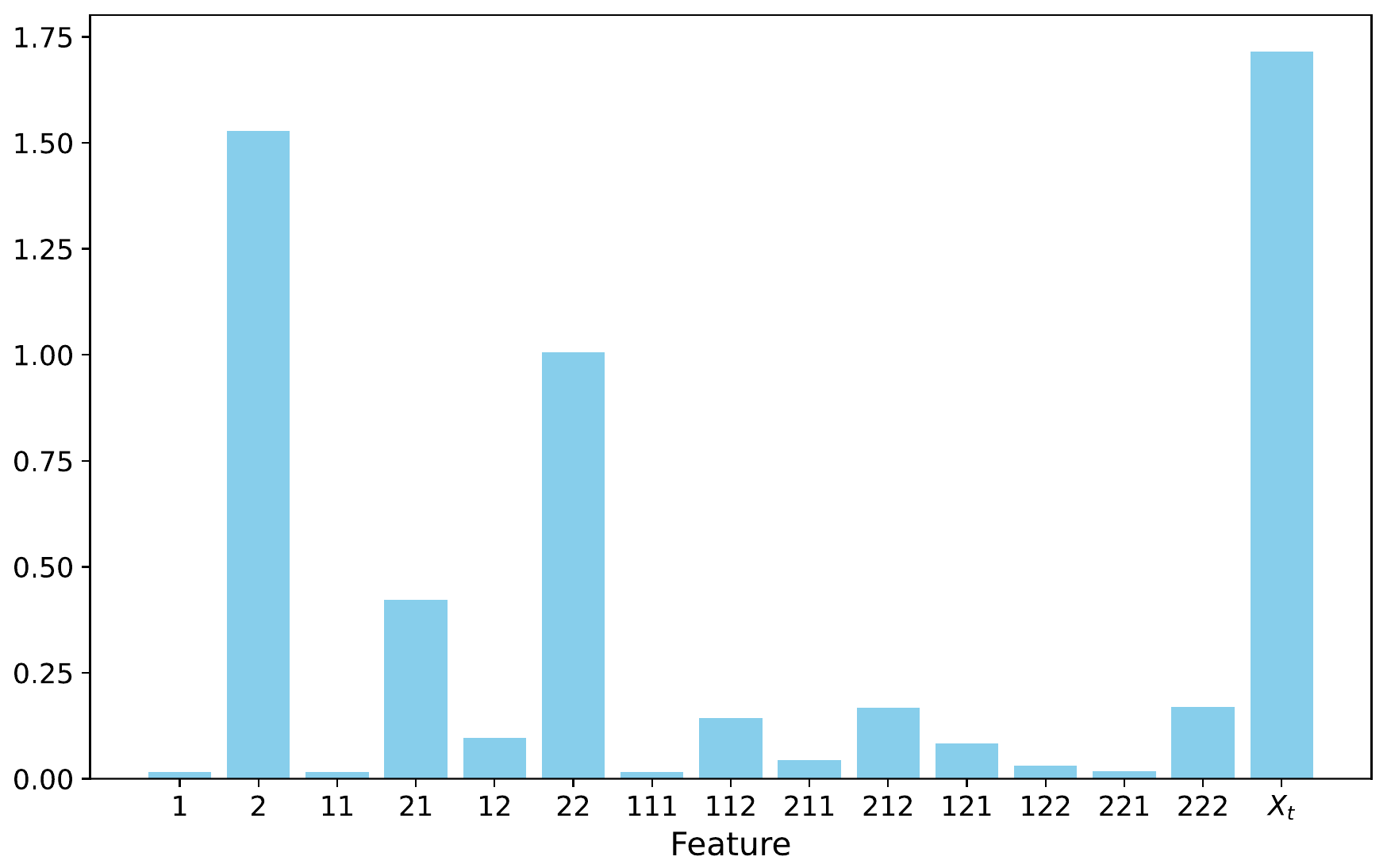}
        \caption{Dual}
        
    \end{subfigure}
    \caption{Feature importance in the rough Bergomi model with $H=0.07$ and $\rho = -0.9$.}
    \label{fig:feature_importance}
\end{figure}

In Figure \ref{fig:discretization} we compare the duality gap in the deep-signature method, with respect to the number of discretization points and different Hurst regimes. More precisely, let us define the relative duality gap \[ 
\epsilon=\epsilon(N,H)= \frac{y_0^{D}(N,H)-y_0^{P}(N,H)}{y_0^{D}(N,H)},\]
where $y_0^{D}(N,H)$ (resp. $y_0^{P}(N,H)$) is the upper (resp. lower) bound obtained from the deep-signature method in the rough Bergomi model with Hurst parameter $H\in (0,1)$ and $N\in\mathbb{N}$.   All the other parameters are chosen the same as in the Tables \ref{tab:putIntervals_08_bergomi}-\ref{tab:putIntervals_H07_Bergomi}. We can observe that in general higher Hurst parameters show smaller duality gaps and also faster convergence with respect to $N$. Notice that, apart from optimization and Monte-Carlo errors, the two main sources for the gap here are the discretization error from approximating iterated and stochastic integrals, and the non-Markovianity for $H\neq 1/2$ regimes. When $H$ is close to $1/2$, the convergence rate seems to be around $H$, and since this regime is "almost Markovian" (and Markovian for $H=1/2$), this can be interpreted as the strong error occuring when approximating the stochastic integrals $\int f(v_t,X_t)dW_t$ with an Euler-scheme and rough integrand. In general, however, it is not possible to seperate these two error sources, so that it becomes more difficult to interpret the observed convergence rate.
Finally, in Figure \ref{fig:feature_importance}  we compare the importance of different entries of the level-$3$ signature in the deep-signature pricing methodologies. This is simply done by first training the primal and dual model, and then for each fixed feature, we shuffle the corresponding samples and measure the error between the  point-estimate and the shuffled-estimate. This procedure we repeat independently $20$ times for each feature and in Figures \ref{fig:feature_importance} we plot the averaged error errors. Notice that the higher this value is, the more influence the corresponding feature has for computing the options price. In both methods, we observe that the "Markovian" state features $X_t$, $v_t =\ip{2}{\hat{\mathbb{V}}_{0,t}}$, $v_t^2=\ip{22}{\hat{\mathbb{V}}_{0,t}}$ appear to have the biggest influence. Apart from that, we can observe that the integrals $\ip{21}{\hat{\mathbb{V}}_{0,t}}=\int_0^tv_ud\langle X \rangle_u$ (resp. $\int_0^tv_udu$), and $\ip{212}{\hat{\mathbb{V}}_{0,t}}=\int_0^t\int_0^uv_rd\langle X \rangle_rdX_u$ (resp. $\ip{212}{\hat{\mathbb{V}}_{0,t}}=\int_0^t\int_0^uv_rdrdX_u$) seem to carry the most important information about the past of the volatility.

\appendix \normalsize
\section{Proof deep-signature optimal stopping}\label{sec:appendix}

\begin{proof}{\textbf{of Proposition \ref{prop:deep}}} The first step is to generalize the global approximation result \cite[Theorem 2.8]{bayer2023primal}. Leaving details to \cite[Section 2]{bayer2023primal}, we are given a measure space $(\mathcal{X},\mathcal{B}(\mathcal{X}),\mu)$, where $\mathcal{X}$ is the space of stopped, time-augmented, geometric $\alpha$-Hölder rough paths (\cite[Definition 2.1]{bayer2023primal}), $\mathcal{B}(\mathcal{X})$ the Borel $\sigma$-algebra, and $\mu$ a measure fulfilling certain assumptions (\cite[Assumption 2.6]{bayer2023primal}). Thanks to \cite[Theorem 2.8]{bayer2023primal}, for any $f\in L^p(\mathcal{X},\mu)$ and $\epsilon>0$, we can find a linear functional on the signature $f^{\epsilon}(\mathbf{x}|_{[0,t]})= \ip{l^\epsilon}{\mathbbm{x}^{<\infty}_{0,t}}$, such that $\Vert f-f_n\Vert_{L^p}\leq \epsilon/2$. But due to $L^p$-universality of DNNs, see \cite[Proposition 2]{leshno1993multilayer}, denoting by $k_{\epsilon}$ the truncation level of $l^{\epsilon}$ in $\mathcal{W}$, for $m$ large enough we can find $(\beta,\tilde{\ell}) \in \mathbb{R}^{m+1} \times (\mathcal{W}^{\leq k_{\epsilon}})^m$ such that $\Vert \theta^{(\beta,\tilde{\ell})}(\mathbf{x}|_{[0,t]}) - f^\epsilon \Vert_{L^p}\leq \epsilon/2$. An application of the triangle inequality then proves that for any $f\in L^p(\mathcal{X},\mu)$, we can find a sequence $(\theta_n)_{n\in \mathbb{N}} \subseteq \mathrm{DNN}^{\sigma}_{\mathrm{sig}}$ such that $\Vert f-\theta_n \Vert_{L^p} \rightarrow 0$ as $n\to \infty$.

Now in the primal case, it is in fact possible to show that for all $0\leq n \leq N$, we have $\mathbb{E}[Z_{\tau_n^{(m,k)}}|\mathcal{F}_{t_n}] \rightarrow \mathbb{E}[Z_{\tau_n}|\mathcal{F}_{t_n}]$ as $(m,k) \to \infty$ in $L^2$, from which the result follows immediately. Using backward induction, one can notice that the result clearly holds true for $n=N$. Assuming that the claim holds for $n+1$, { we denote by $\theta^{(m,k)}$ the minimizer in \eqref{eq:deep_primal_minimizer}. } We can estimate $ \Vert \mathbb{E}[Z_{\tau_n^{(m,k)}}-Z_{\tau_n}|\mathcal{F}_{t_n}] \Vert_{L^2}$ exactly as it was done \cite[Appendix A.1]{bayer2023primal} until equation (A.1), to reduce the claim to the convergence of \begin{equation*}
    L^{(m,k)}=\big \Vert \theta^{(m,k)}(\mathbf{X}|_{[0,t_n]})-\mathbb{E}[Z_{\tau_{n+1}}|\mathcal{F}_{t_n}] \big \Vert_{L^2}.
\end{equation*} There is, however, a subtle difference here compared to the conclusion used in the linear case: { The sets $\mathcal{D}_\eta^{(m,k)}$, and in particular also $\mathrm{DNN}^{\sigma}_{\mathrm{sig}}$, are not convex,} which means that $\theta^{(m,k)}(\mathbf{X}|_{[0,t_n]})$ with respect to \eqref{eq:deep_primal_minimizer} is not necessary the orthogonal projection of $Z_{\tau_{n+1}^{(k,m)}}$. We can still conclude by introducing the minimizer { $\hat{\theta}^{(m,k)}$ } similar to \eqref{eq:deep_primal_minimizer}, but replacing $Z_{\tau_{n+1}^{(k,m)}}$ by $\mathbb{E}[Z_{\tau_{n+1}}|\mathcal{F}_{t_n}]$. Then we have \begin{align*}
    L^{(m,k)} & \leq \big \Vert \theta^{(m,k)}(\mathbf{X}|_{[0,t_n]})-\mathbb{E}[Z_{\tau^{(m,k)}_{n+1}}|\mathcal{F}_{t_n}] \big \Vert_{L^2} +\big \Vert \mathbb{E}[Z_{\tau^{(m,k)}_{n+1}}-Z_{\tau_{n+1}}|\mathcal{F}_{t_n}] \big \Vert_{L^2} \\ & \leq \big \Vert \hat{\theta}^{(m,k)}(\mathbf{X}|_{[0,t_n]})-\mathbb{E}[Z_{\tau^{(m,k)}_{n+1}}|\mathcal{F}_{t_n}] \big \Vert_{L^2} + \big \Vert \mathbb{E}[Z_{\tau^{(m,k)}_{n+1}}-Z_{\tau_{n+1}}|\mathcal{F}_{t_n}] \big \Vert_{L^2} \\ & \leq \big \Vert \hat{\theta}^{(m,k)}(\mathbf{X}|_{[0,t_n]})-\mathbb{E}[Z_{\tau_{n+1}}|\mathcal{F}_{t_n}] \big \Vert_{L^2}+2\big \Vert \mathbb{E}[Z_{\tau^{(m,k)}_{n+1}}-Z_{\tau_{n+1}}|\mathcal{F}_{t_n}] \big \Vert_{L^2} \\ &  \leq \big \Vert \hat{\theta}^{(m,k)}(\mathbf{X}|_{[0,t_n]})-\mathbb{E}[Z_{\tau_{n+1}}|\mathcal{F}_{t_n}] \big \Vert_{L^2} + 2\big \Vert \mathbb{E}[Z_{\tau^{(m,k)}_{n+1}}-Z_{\tau_{n+1}}|\mathcal{F}_{t_{n+1}}] \big \Vert_{L^2},
\end{align*} where we used the triangle inequality in the first and third inequality, the fact that $\theta^{(m,k)}$ minimizes the $L^2$ distance to $\mathbb{E}[Z_{\tau^{(m,k)}_{n+1}}|\mathcal{F}_{t_n}]$ in the second inequality, and the contraction property of conditional expectations in the last inequality. Now the last term in the inequality converges by induction hypothesis, while the first one converges thanks to the global approximation result as $m,k \to \infty$.

Finally, let us outline also the proof of the dual case, which closely follows the techniques used in \cite[Appendix A.2]{bayer2023primal}. First, an application of the global approximation allows us to equivalently write \eqref{eq:dual_optimalstopping} as \begin{equation}\label{eq:dual_equivalence}
    Y^N_0 = \inf_{\theta \in \mathrm{DNN}^{\sigma}_{\mathrm{sig}}}\mathbb{E}\big [\max_{0\leq n \leq N}(Z_{t_n}-M^{\theta}_{t_n})\big ], \quad M^{\theta}_{t_n} = \int_0^{t_n} \theta(\mathbf{X}|_{[0,s]})dW_s
\end{equation} Indeed, leaving the detailed technique to the proof of \cite[Theorem 3.7]{bayer2023primal}, this follows from the following observation: As already discussed in the beginning of Section \ref{sec:dual_OSP}, since $\mathcal{F}^X=\mathcal{F}^W$, by martingale representation it is enough to minimize \eqref{eq:dual_optimalstopping} over $L^2$-progressive processes, which in turn can by approximated arbitrary well by $\theta \in \mathrm{DNN}^{\sigma}_{\mathrm{sig}}$ thanks to the global approximation result. Next we define $Y_0^{N,(m,k)} = \mathbb{E}[\max_{0 \leq n \leq N}(Z_{t_n}-M^{(m,k)}_{t_n})]$, where we recall $M^{(m,k)}$ was defined with respect to the minimizer \eqref{eq:deep_dual_minimizer}, and note that $Y_0^{N,(m,k)}\geq Y_0^N$. Then, by \eqref{eq:dual_equivalence}, for every $\epsilon>0$, there exists an element $\theta^{\epsilon} \in \mathrm{DNN}^{\sigma}_{\mathrm{sig}}$ such that $Y_0^N+\epsilon \geq \mathbb{E}[\max_{0 \leq n \leq N}(Z_{t_n}-M^{\theta^\epsilon}_{t_n})]$, so that \[
0 \leq Y_0^{N,(m,k)}-Y^N_0 \leq \epsilon + Y_0^{N,(m,k)}-\mathbb{E}\big [\max_{0 \leq n \leq N}(Z_{t_n}-M^{\theta^\epsilon}_{t_n})\big ] \leq \epsilon,
\] where the last inequality holds for $(m,k)$ large enough, such that $\theta^\epsilon=\theta^{(\tilde{\beta},\tilde{\ell})}$ for some $(\tilde{\beta},\tilde{\ell}) \in \mathbb{R}^m \times (\mathcal{W}^{\leq k})^m$ and $\Vert \theta^{(\tilde{\beta},\tilde{\ell})} \Vert \leq \eta_m$.

\end{proof} 

\begin{remark}\label{rmk:log_sig_rmk}
    The same results remain true when working with the log-signature introduced in Remark \ref{rmk:log-signature}. For the global approximation result, one can simply note that for any $f\in L^p(\mathcal{X},\mu)$, we know that $f \overset{L^p}{\approx}\ip{\ell}{\mathbb{X}^{\leq K}}=\ip{\ell}{\mathrm{exp}^\otimes(\mathbbm{L}^{\leq K})}  $ for $K$ large enough, where $\exp^{\otimes}$ is the (continuous!) inverse of the $\mathrm{log}^\otimes$ introduced in Remark \ref{rmk:log-signature}. Using again \cite[Proposition 2]{leshno1993multilayer}, we can approximate this exponential on the truncated log-signature by a DNN. In summary, $f \overset{L^p}{\approx} \sum_j \beta_j \sigma(\ip{\tilde{\ell_j}}{\mathbb{L}^{<\infty}})$ in the sense of the global approximation result discussed in the beginning of the proof. Then, all the arguments can be repeated, relying on the set $\mathrm{DNN}_{\mathrm{log}}^\sigma$, which replaces the signature with the log-signature.
\end{remark}
\bibliographystyle{alpha}
\bibliography{biblio}
\end{document}